\documentclass[12pt]{article}

\usepackage{amssymb}
\usepackage{amsthm}

\usepackage{logique}

\usepackage{graphicx} 

\renewcommand{\force}{\;\|\!\!\!-\,}
\renewcommand{\vv}{|\hspace{-0.1em}|\hspace{-0.1em}|}
\renewcommand{\fforce}{\;|\hspace{-0.1em}|\hspace{-0.1em}|\hspace{-0.45em}-\,}
\renewcommand{\models}{\;|\!\!\!=\,}
\renewcommand{\succc}{\succ\!\!\!\succ}

\newcommand{\NN}{\mathbb{N}}

\newcommand{\OO}{\mathbb{O}}

\newcommand{\RR}{\mathbb{R}}

\newcommand{\BBB}{\mbox{\sffamily B}}
\newcommand{\CCC}{\mbox{\sffamily C}}

\newcommand{\III}{\mbox{\sffamily I}}
\newcommand{\HHH}{\mbox{\sffamily H}}
\newcommand{\KKK}{\mbox{\sffamily K}}
\newcommand{\WWW}{\mbox{\sffamily W}}
\newcommand{\PP}{\mbox{\footnotesize\sffamily P}}
\newcommand{\dd}{\mbox{\sffamily d}}

\newcommand{\kkk}{\mbox{\bf\sffamily k}}
\newcommand{\cccc}{\mbox{\bf\sffamily cc}}

\newcommand{\indi}{^{\mbox{\footnotesize int}}}

\newcommand{\trgl}{\triangleleft}

\newcommand{\bptho}{\mbox{\bfseries(\hspace{-0.4em}(}}
\newcommand{\bpthf}{\mbox{\bfseries)\hspace{-0.4em})}}
\newcommand{\bpptho}{\mbox{\large\bfseries(\hspace{-0.4em}(}}
\newcommand{\bppthf}{\mbox{\large\bfseries)\hspace{-0.4em})}}
\newcommand{\bppptho}{\mbox{\Large\bfseries(\hspace{-0.4em}(}}
\newcommand{\bpppthf}{\mbox{\Large\bfseries)\hspace{-0.4em})}}

\newcommand{\et}{{\scriptstyle\land}}
\newcommand{\ou}{{\scriptstyle\lor}}

\newcommand{\lle}{\,{\scriptstyle\ll}\,}
\newcommand{\mlbd}{\reflectbox{$\lambda$}}

\newcommand{\ZFe}{ZF$_\varepsilon$}

\newcommand{\hto}{\hookrightarrow}

\newtheorem{theorem}{Theorem}
\newtheorem{lemma}[theorem]{Lemma}
\newtheorem{corollary}[theorem]{Corollary}
\newtheorem{proposition}[theorem]{Proposition}

\author{Jean-Louis Krivine\\
{\small University Paris-Diderot - CNRS}}

\title{Realizability algebras III~: some examples}
\date{\footnotesize {\today}}

\begin{document}
\maketitle\noindent
\section*{Introduction}\noindent
The notion of \emph{realizability algebra}, which was introduced in \cite{krivine5,krivine6},
is a tool to study the proof-program correspondence and to build new models of set theory, which we call
\emph{realizability models of ZF}.\\
It is a variant of the well known notion of \emph{combinatory algebra},
with a new instruction~$\ccc$, and a new type for the \emph{environments}.\\
The \emph{sets of forcing conditions}, in common use in set theory, are (very) particular cases
of realizability algebras~; and the forcing models of ZF are very particular cases of realizability models.

\smallskip\noindent
We show here how to extend an arbitrary realizability algebra, by means of a certain set of conditions, so that the axiom DC of dependent choice is realized.\\
In order to avoid introducing  new instructions, we use an idea of A.~Miquel~\cite{miquel1}.\\
This technique has applications of two kinds~:

\smallskip\noindent
1. Construction of models of ZF + DC.\\
When the initial realizability algebra is not trivial (that is, if we are not in the case of forcing or
equivalently, if the associated Boolean algebra $\gimel2$ is $\ne\{0,1\}$), then we always obtain in this way
a model of ZF \emph{which satisfies DC + there is no well ordering of \ $\RR$}.\\
By suitably choosing the realizability algebras, we can get, for instance, the relative consistency
over ZF of the following two theories~:

\smallskip\noindent
i)~~ZF + DC + there exists an increasing function $i\mapsto X_i$, from the countable atomless
Boolean algebra ${\cal B}$ into ${\cal P}(\RR)$ such that~:\\
$X_0=\{0\}$~; $i\ne0$ $\Fl$ $X_i$ is uncountable~;\\
$X_i\cap X_j=X_{i\land j}$~;\\
if \ $i\land j=0$ \ then \ $X_{i\lor j}$ is equipotent with $X_i\fois X_j$~;\\
$X_i\fois X_i$ is equipotent with $X_i$~;\\
there exists a surjection from $X_1$ onto $\RR$~;\\
if there exists a surjection from $X_j$ onto $X_i$, then $i\le j$~;\\
if \ $i,j\ne0,i\land j=0$, there is no surjection from \ $X_i\oplus X_j$ onto $X_i\fois X_j$~;\\
more generally, if $A\subset{\cal B}$ and if there exists a surjection from \
$\bigcup_{j\in A}X_j$ onto $X_i$, then $i\le j$ for some $j\in A$.

\smallskip\noindent
In particular, there exists a sequence of subsets of $\RR$, the cardinals of which are not comparable, and also
a sequence of subsets of $\RR$, the cardinals of which are strictly decreasing.
 
\smallskip\noindent
ii)~~ZF + DC + \ there exists $X\subset\RR$ such that~:\\
$X$ is uncountable and there is no surjection from $X$ onto $\aleph_1$\\
(and therefore, every well orderable subset of $X$ is countable)~;\\
$X\fois X$ is equipotent with $X$~;\\
there exists a total order on $X$, every proper initial segment of which is countable~;\\
there exists a surjection from $X\fois\aleph_1$ onto $\RR$~;\\
there exists an injection from $\aleph_1$ (thus also from $X\fois\aleph_1$) into $\RR$.

\smallskip\noindent
2.~~Curry-Howard correspondence.\\
With this technique of extension of realizability algebras, we can obtain a program from a proof,
in \ ZF + DC, of an arithmetical formula $F$, which is a $\lbd_c$-term, that is, a $\lbd$-term
containing $\ccc$, \emph{but no other new instruction}.\\
This is a notable difference with the method given in \cite{krivine2,krivine3}, where we use
the instruction {\tt quote} and which is, on the other hand, simpler and not limited to
arithmetical formulas.

\smallskip\noindent
It is important to observe that the program we get in this way does not really depend on the given
proof of \ $DC\to F$ \ in \ ZF\/, but only on the \emph{program P extracted from this proof},
which is a closed $\lbd_c$-term.
Indeed, we obtain this program by means of an operation of \emph{compilation} applied to~P
(look at the remark at the end of the introduction of~\cite{krivine5}).

\smallskip\noindent
Finally,  apart from applications 1 and 2, we may notice theorem~\ref{kappa+nbo}, which gives
an interesting property of \emph{every} realizability model~: as soon as the Boolean algebra~$\gimel2$
is not trivial (i.e. if the model is not a forcing model), there exists a non well orderable individual.

\section{Generalities}\label{general}\noindent
\subsection*{Realizability algebras}\noindent
It is a first order structure, which is defined in~\cite{krivine5}. We recall here briefly
the definition and some essential properties~:

\smallskip\noindent
A \emph{realizability algebra} ${\cal A}$ is made up of three sets~:
$\Lbd$ (the set of \emph{terms}), $\Pi$ (the set of \emph{stacks}),
$\Lbd\star\Pi$ (the set of \emph{processes}) with the following operations~:

\smallskip\noindent
$(\xi,\eta)\mapsto(\xi)\eta$ from $\Lbd^2$ into $\Lbd$ (\emph{application})~;\\
$(\xi,\pi)\mapsto\xi\ps\pi$ from $\Lbd\fois\Pi$ into $\Pi$ (\emph{push})~;\\
$(\xi,\pi)\mapsto\xi\star\pi$ from $\Lbd\fois\Pi$ into $\Lbd\star\Pi$ (\emph{process})~;\\
$\pi\mapsto\kk_\pi$ from $\Pi$ into $\Lbd$ (\emph{continuation}).

\smallskip\noindent
There are, in $\Lbd$, distinguished elements \ $\BBB,\CCC,\III,\KKK,\WWW,\ccc$, called
\emph{elementary combinators} or \emph{instructions}.

\smallskip\noindent
{\bfseries Notation.}\\
The term $(\ldots(((\xi)\eta_1)\eta_2)\ldots)\eta_n$ will be also written as
$(\xi)\eta_1\eta_2\ldots\eta_n$ or $\xi\eta_1\eta_2\ldots\eta_n$.\\
For instance~: \ $\xi\eta\zeta=(\xi)\eta\zeta=(\xi\eta)\zeta=((\xi)\eta)\zeta$. 

\smallskip\noindent
We define a preorder on $\Lbd\star\Pi$, denoted by \ $\succ$, which is called \emph{execution}~;\\ $\xi\star\pi\succ\xi'\star\pi'$ is read as~:
\emph{the process $\xi\star\pi$ reduces to $\xi'\star\pi'$.}\\
It is the smallest reflexive and transitive binary relation, such that, for any $\xi,\eta,\zeta\in\Lbd$ and $\pi,\varpi\in\Pi$, we have~:

\smallskip\noindent
$(\xi)\eta\star\pi\succ\xi\star\eta\ps\pi$.\\
$\III\star\xi\ps\pi\succ\xi\star\pi$.\\
$\KKK\star\xi\ps\eta\ps\pi\succ\xi\star\pi$.\\
$\WWW\star\xi\ps\eta\ps\pi\succ\xi\star\eta\ps\eta\ps\pi$.\\
$\CCC\star\xi\ps\eta\ps\zeta\ps\pi\succ\xi\star\zeta\ps\eta\ps\pi$.\\
$\BBB\star\xi\ps\eta\ps\zeta\ps\pi\succ\xi\star(\eta)\zeta\ps\pi$.\\
$\ccc\star\xi\ps\pi\succ\xi\star\kk_\pi\ps\pi$.\\
$\kk_\pi\star\xi\ps\varpi\succ\xi\star\pi$.

\smallskip\noindent
We are also given a subset $\bbot$ of $\Lbd\star\Pi$ such that~:\\
\centerline{$\xi\star\pi\succ\xi'\star\pi'$, $\xi'\star\pi'\in\bbot$ \ $\Fl$ \ $\xi\star\pi\in\bbot$.}

\smallskip\noindent
Given two processes $\xi\star\pi,\xi'\star\pi'$, the notation $\xi\star\pi\succc\xi'\star\pi'$ means~:\\ \centerline{$\xi\star\pi\notin\bbot\Fl\xi'\star\pi'\notin\bbot$.}

\smallskip\noindent
Therefore, obviously, \ $\xi\star\pi\succ\xi'\star\pi'\;\Fl\;\xi\star\pi\succc\xi'\star\pi'$.

\medskip\noindent
Finally, we choose a set of terms \ QP$_{\cal A}\subset\Lbd$, containing the elementary combinators~:\\ $\BBB,\CCC,\III,\KKK,\WWW,\ccc$ and closed by application. They are called the
\emph{proof-like terms of the algebra ${\cal A}$}. We write also \ QP instead of
QP$_{\cal A}$ if there is no ambiguity about ${\cal A}$.\\
The algebra ${\cal A}$ is called \emph{coherent} if, for every proof-like term
$\theta\in\mbox{QP}_{\cal A}$, there exists a stack~$\pi$ such that $\theta\star\pi\notin\bbot$.

\smallskip\noindent
{\small{\bfseries{}Remark.} The \emph{sets of forcing conditions} can be considered as degenerate cases of realizability
algebras, if we present them in the following way~: an inf-semi-lattice $P$, with a greatest element~$1$ and an
initial segment $\bbot$ of $P$ (the set of \emph{false} conditions). Two conditions $p,q\in P$ are called
\emph{compatible} if their g.l.b. $p\et q$ is not in $\bbot$.\\
We get a realizability algebra if we set $\Lbd=\Pi=\Lbd\star\Pi=P$~; $\BBB=\CCC=\III=\KKK=\WWW=\ccc=1$ and QP$=\{1\}$~;
$(p)q=p\ps q=p\star q=p\et q$ and $\kk_p=p$. The preorder $p\succ q$ is defined as $p\le q$, i.e. $p\et q=p$. The condition of coherence is $1\notin\bbot$.}

\subsection*{\cc-terms and $\lbd$-terms}
The terms of the language of combinatory algebra, which are built with variables, elementary combinators and
the application (binary operation), will be called \emph{combinatory terms} or \emph{$\cc$-terms},
in order to distinguish them from the terms of the algebra ${\cal A}$, which are elements of $\Lbd$.\\
Each closed $\cc$-term (i.e. without variable) takes a value in the algebra ${\cal A}$,
which is a proof-like term of ${\cal A}$.

\smallskip\noindent
Let us call \emph{atom} a $\cc$-term of length $1$, i.e. a constant symbol
$\BBB,\CCC,\III,\KKK,\WWW,\ccc$ or a variable.

\begin{lemma}\label{att}
Every $\cc$-term $t$ can be written, in a unique way,
in the form $t=(a)t_1\ldots t_k$ where $a$ is an atom and $t_1,\ldots,t_k$ are $\cc$-terms.
\end{lemma}\noindent
Immediate, by recurrence on the length of $t$.

\qed

\smallskip\noindent
The result of the \emph{substitution of \
$\xi_1,\ldots,\xi_n\in\Lbd$ to the variables $x_1,\ldots,x_n$} in a $\cc$-term $t$,
is a \emph{term} (i.e. an element of $\Lbd$) denoted by $t[\xi_1/x_1,\ldots,\xi_n/x_n]$ or,
more briefly, $t[\vec{\xi}/\vec{x}]$.\\
The inductive definition is~:

\smallskip\noindent
$a[\vec{\xi}/\vec{x}]=\xi_i$ if $a=x_i(1\le i\le n)$~;\\
$a[\vec{\xi}/\vec{x}]=a$ if $a$ is an atom $\ne x_1,\ldots,x_n$~;\\
$(tu)[\vec{\xi}/\vec{x}]=(t[\vec{\xi}/\vec{x}])u[\vec{\xi}/\vec{x}]$.

\smallskip\noindent
Given a \cc-term $t$ and a variable $x$, we define inductively on $t$, a new \cc-term denoted by
$\mlbd x\,t$, which does not contain $x$. To this aim, we apply the first possible case in the following list~:

\smallskip\noindent
1.~$\mlbd x\,t=(\KKK)t$\label{def_lbd} if $t$ does not contain $x$.\\
2.~$\mlbd x\,x=\,\III$.\\
3.~$\mlbd x\,tu=(\CCC\mlbd x\,t)u$ if $u$ does not contain $x$.\\
4.~$\mlbd x\,tx=t$ if $t$ does not contain $x$.\\
5.~$\mlbd x\,tx=(\WWW)\mlbd x\,t$ (if $t$ contains $x$).\\
6.~$\mlbd x(t)(u)v=\mlbd x(\BBB)tuv$ (if $uv$ contains $x$).

\smallskip\noindent
It is easy to see that this rewriting is finite, for any given $\cc$-term $t$~: indeed, during the rewriting, no combinator is introduced inside $t$, but only in front of it. Moreover, the only changes in~$t$ are~: moving
parentheses and erasing occurrences of $x$. Now, rules 1~to~5 strictly decrease, and rule~6 does not increase,
the part of $t$ which remains under $\mlbd x$. Moreover, rule~6 can be applied consecutively only finitely many times.

\smallskip\noindent
Given a \cc-term $t$ and a variable $x$, we now define the $\cc$-term $\lbd x\,t$ by setting~:\\
$\lbd x\,t=\mlbd x\,(\III)t$.

\smallskip\noindent
This enables us to translate every $\lbd$-term into a $\cc$-term.
In the sequel, almost all $\cc$-terms will be written as $\lbd$-terms.

\smallskip\noindent
The fundamental property of this translation is given by theorem~\ref{beta_red_gauche}~:

\begin{theorem}\label{beta_red_gauche}
Let $t$ be a \ \cc-term with the only variables $x_1,\ldots,x_n$~;
let \ $\xi_1,\ldots,\xi_n\in\Lbd$ and $\pi\in\Pi$. Then
$\lbd x_1\ldots\lbd x_n\,t\star\xi_1\ps\ldots\ps\xi_n\ps\pi
\succ t[\xi_1/x_1,\ldots,\xi_n/x_n]\star\pi$.
\end{theorem}\noindent

\begin{lemma}\label{brg1}
Let $a$ be an atom, $t=(a)t_1\ldots t_k$ a \cc-term with the only variables $x,y_1,\ldots,y_n$,  and \
$\xi,\eta_1,\ldots,\eta_n\in\LLbd$~; then~:\\
$(\mlbd x\,t)[\vec{\eta}/\vec{y}]\star\xi\ps\pi\succ a[\xi/x,\vec{\eta}/\vec{y}]\star
t_1[\xi/x,\vec{\eta}/\vec{y}]\ps\ldots\ps t_k[\xi/x,\vec{\eta}/\vec{y}]\ps\pi$.
\end{lemma}\noindent
The proof is done by induction on the number of rules~1 to~6 used to translate the term $\mlbd x\,t$.
Consider the rule used first.

\smallskip\noindent
$\bullet$~Rule~1~: we have $(\mlbd x\,t)[\vec{\eta}/\vec{y}]\star\xi\ps\pi
\equiv(\KKK)t[\vec{\eta}/\vec{y}]\star\xi\ps\pi
\succ\KKK\star t[\vec{\eta}/\vec{y}]\ps\xi\ps\pi\succ t[\vec{\eta}/\vec{y}]\star\pi\\
\equiv t[\xi/x,\vec{\eta}/\vec{y}]\star\pi$ because $x$ is not in $t$.
The result follows immediately.

\smallskip\noindent
$\bullet$~Rule~2~: we have $t=x,\mlbd x\,t=\III$ and the result is trivial.

\smallskip\noindent
In rules~3, 4, 5 or 6, we have $t=ut_k$ with $u=at_1\ldots t_{k-1}$, by lemma~\ref{att}.

\smallskip\noindent
$\bullet$~Rule~3~: $(\mlbd x\,t)[\vec{\eta}/\vec{y}]\star\xi\ps\pi\equiv
((\CCC\mlbd x\,u)t_k)[\vec{\eta}/\vec{y}]\star\xi\ps\pi
\succ\CCC\star(\mlbd x\,u)[\vec{\eta}/\vec{y}]\ps t_k[\vec{\eta}/\vec{y}]\ps\xi\ps\pi\\
\succ(\mlbd x\,u)[\vec{\eta}/\vec{y}]\star\xi\ps t_k[\vec{\eta}/\vec{y}]\ps\pi\\
\succ a[\xi/x,\vec{\eta}/\vec{y}]\star
t_1[\xi/x,\vec{\eta}/\vec{y}]\ps\ldots\ps t_{k-1}[\xi/x,\vec{\eta}/\vec{y}]\ps t_k[\vec{\eta}/\vec{y}]\pi$
by the induction hypothesis\\
$\equiv a[\xi/x,\vec{\eta}/\vec{y}]\star
t_1[\xi/x,\vec{\eta}/\vec{y}]\ps\ldots\ps t_{k-1}[\xi/x,\vec{\eta}/\vec{y}]\ps t_k[\xi/x,\vec{\eta}/\vec{y}]\ps\pi$
since $x$ is not in~$t_k$.

\smallskip\noindent
In rules 4 and 5, we have $t_k=x$, i.e. $t=(u)x$.

\smallskip\noindent
$\bullet$~Rule~4~: we have $(\mlbd x\,t)[\vec{\eta}/\vec{y}]\star\xi\ps\pi\equiv
u[\vec{\eta}/\vec{y}]\star\xi\ps\pi\equiv u[\xi/x,\vec{\eta}/\vec{y}]\star\xi\ps\pi$
because $x$ is not in $u$. Since $u=at_1\ldots t_{k-1}$ and $t_k=x$, the result follows immediately.

\smallskip\noindent
$\bullet$~~Rule~5~: we have $t_k=x$ and $(\mlbd x\,t)[\vec{\eta}/\vec{y}]\star\xi\ps\pi\equiv
(\WWW\mlbd x\,u)[\vec{\eta}/\vec{y}]\star\xi\ps\pi\\
\succ\WWW\star(\mlbd x\,u)[\vec{\eta}/\vec{y}]\ps\xi\ps\pi
\succ(\mlbd x\,u)[\vec{\eta}/\vec{y}]\star\xi\ps\xi\ps\pi\\
\succ a[\xi/x,\vec{\eta}/\vec{y}]\star
t_1[\xi/x,\vec{\eta}/\vec{y}]\ps\ldots\ps t_{k-1}[\xi/x,\vec{\eta}/\vec{y}]\ps\xi\ps\pi$
(by the induction hypothesis)\\
$\equiv a[\xi/x,\vec{\eta}/\vec{y}]\star
t_1[\xi/x,\vec{\eta}/\vec{y}]\ps\ldots\ps t_k[\xi/x,\vec{\eta}/\vec{y}]\ps\pi$.

\smallskip\noindent
$\bullet$~~Rule~6~: we have $t_k=(v)w$ \ and \ $(\mlbd x\,t)[\vec{\eta}/\vec{y}]\star\xi\ps\pi\equiv
(\mlbd x(\BBB)uvw)[\vec{\eta}/\vec{y}]\star\xi\ps\pi\\
\succ\BBB\star u[\xi/x,\vec{\eta}/\vec{y}]\ps v[\xi/x,\vec{\eta}/\vec{y}]\ps w[\xi/x,\vec{\eta}/\vec{y}]\ps\pi$
(by the induction hypothesis)\\
$\succ u[\xi/x,\vec{\eta}/\vec{y}]\star t_k[\xi/x,\vec{\eta}/\vec{y}]\ps\pi\\
\succ a[\xi/x,\vec{\eta}/\vec{y}]\star
t_1[\xi/x,\vec{\eta}/\vec{y}]\ps\ldots\ps t_{k-1}[\xi/x,\vec{\eta}/\vec{y}]
\ps t_k[\xi/x,\vec{\eta}/\vec{y}]\ps\pi$.

\qed

\begin{lemma}\label{brg2}
$(\lbd x\,t)[\vec{\eta}/\vec{y}]\star\xi\ps\pi\succ t[\xi/x,\vec{\eta}/\vec{y}]\star\pi$.
\end{lemma}\noindent
Immediate  by lemma~\ref{brg1} and the definition of $\lbd x\,t$ which is $\mlbd x(\III)t$.

\qed

\smallskip\noindent
We can now prove theorem~\ref{beta_red_gauche} by induction on $n$~; the case $n=0$ is trivial.\\
We have \ $\lbd x_1\ldots\lbd x_{n-1}\lbd x_n\,t\star\xi_1\ps\ldots\ps\xi_{n-1}\ps\xi_n\ps\pi\succ
(\lbd x_nt)[\xi_1/x_1,\ldots,\xi_{n-1}/x_{n-1}]\star\xi_n\ps\pi$\\
(by induction hypothesis) \ $\succ t[\xi_1/x_1,\ldots,\xi_{n-1}/x_{n-1},\xi_n/x_n]\star\pi$
by lemma~\ref{brg2}.

\qed

\subsection*{The formal system}
We write formulas and proofs in the language of first order logic. This formal
language consists of~:

\smallskip\noindent
$\bullet$~~\emph{individual variables} \ $x,y,\ldots$~;\\
$\bullet$~~\emph{function symbols} $f,g,\ldots$ of various arities~; function symbols of arity~$0$ are called \emph{constant symbols}.\\
$\bullet$~~\emph{relation symbols}~; there are three binary relation symbols~: $\neps,\notin,\subset$.

\smallskip\noindent
The terms of this first order language will be called \emph{$\ell$-terms}~; they are built in the usual way with individual variables and function symbols.

\smallskip\noindent
{\small{\bfseries Remark.} Thus, we use four expressions with the word \emph{term}~: term, $\cc$-term,
$\lbd$-term and $\ell$-term.}

\smallskip\noindent
The \emph{atomic formulas} are the expressions $\top,\bot,t\neps u,t\notin u,t\subset u$,
where $t,u$ are $\ell$-terms.

\smallskip\noindent
\emph{Formulas} are built as usual, from atomic formulas, \emph{with the only logical symbols} \
$\to$, $\pt$~:\\
$\bullet$~~each atomic formula is a formula~;\\
$\bullet$~~if $A,B$ are formulas, then $A\to B$ is a formula~;\\
$\bullet$~~if $A$ is a formula and $x$ an individual variable, then $\pt x\,A$ is a
formula.

\smallskip\noindent
{\bfseries Notations.} Let $A_1,\ldots,A_n,A,B$ be formulas. Then~:\\
$A\to\bot$ \ is written \ $\neg A$~;\\
$A_1\to(A_2\to\cdots\to(A_n\to B)\cdots)$ \ is written \ $A_1,A_2,\ldots,A_n\to B$~;\\
$\neg A_1,\ldots,\neg A_n\to\bot$ \ is written \ $A_1\lor\ldots\lor A_n$~;\\
$(A_1,\ldots,A_n\to\bot)\to\bot$ \ is written \ $A_1\land\ldots\land A_n$~;\\
$\neg\pt x(A_1,\ldots,A_n\to\bot)$ \ is written \ $\ex x\{A_1,\ldots,A_n\}$.

\smallskip\noindent
The \emph{rules of natural deduction} are the following (the $A_i$'s are formulas, the $x_i$'s
are variables of \cc-term, $t,u$ are \cc-terms, written as $\lbd$-terms)~:

\smallskip\noindent
1.~$x_1:A_1,\ldots,x_n:A_n\vdash x_i:A_i$.\\
2.~$x_1:A_1,\ldots,x_n:A_n\vdash t:A\to B$, \ \ $x_1:A_1,\ldots,x_n:A_n\vdash u:A$ \ \
$\Fl$ \ \ $x_1:A_1,\ldots,x_n:A_n\vdash tu:B$.\\
3.~$x_1:A_1,\ldots,x_n:A_n,x:A\vdash t:B$ \ \ $\Fl$ \ \
$x_1:A_1,\ldots,x_n:A_n\vdash\lbd x\,t:A\to B$.\\
4.~$x_1:A_1,\ldots,x_n:A_n\vdash t:A$ \ \ $\Fl$ \ \
$x_1:A_1,\ldots,x_n:A_n\vdash t:\pt x\,A$ \ \
where $x$ is an individual variable which does not appear in $A_1,\ldots,A_n$.\\
5.~$x_1:A_1,\ldots,x_n:A_n\vdash t:\pt x\,A$ \ \ $\Fl$ \ \
$x_1:A_1,\ldots,x_n:A_n\vdash t:A[\tau/x]$ \ \ where $x$ is an individual variable and
$\tau$ is a $\ell$-term.\\
6.~$x_1:A_1,\ldots,x_n:A_n\vdash\ccc:((A\to B)\to A)\to A$ \ (law of Peirce).\\
7.~$x_1:A_1,\ldots,x_n:A_n\vdash t:\bot$ \ \ $\Fl$ \ \
$x_1:A_1,\ldots,x_n:A_n\vdash t:A$ \ \ for every formula $A$.

\subsection*{Realizability models}\noindent
We formalize set theory with the first order language described above. We write, in this language,
the axioms of a theory named \ZFe, which are given in \cite{krivine6}.\\
The usual set theory ZF is supposed written with the only relation symbols $\notin,\subset$.\\
Then, \ZFe\  is a \emph{conservative extension} of ZF\/, which is proved in \cite{krivine6}.

\smallskip\noindent
Let us consider a  \emph{coherent} realizability algebra ${\cal A}$, defined in
a model ${\cal M}$ of ZFL, which is called the \emph{ground model}. The elements
of ${\cal M}$ will be called \emph{individuals} (in order to avoid the word \emph{set},
as far as possible).

\smallskip\noindent
We defined, in~\cite{krivine6}, a \emph{realizability model}, denoted\, by
${\cal N}_{\cal A}$ (or even ${\cal N}$, if there is no ambiguity about the algebra ${\cal A}$).\\
It has the same domain (the same individuals) as ${\cal M}$ and the interpretation of the function symbols
is the same as in ${\cal M}$.

\smallskip\noindent
Each closed formula $F$ of \ZFe\ with parameters in ${\cal M}$, has \emph{two truth values} in ${\cal N}$,
which are denoted by $\|F\|$ (which is a subset of $\Pi$) and $|F|$ (which is a subset of $\Lbd$).\\
Here are their definitions~:

\smallskip\noindent
$|F|$ is defined immediately from $\|F\|$ \ as follows~:\\
\centerline{$\xi\in|F|$ \ $\Dbfl$ \ $(\pt\pi\in\|F\|)\,\xi\star\pi\in\bbot$.}

\smallskip\noindent
We shall write \ $\xi\force F$ \ (read \emph{``~$\xi$ realizes $F$~''}) for \ $\xi\in|F|$.

\smallskip\noindent
$\|F\|$ is now defined by recurrence on the length of $F$~:

\smallskip\noindent
$\bullet$~~$F$ is atomic~;\\
then $F$ has one of the forms \ $\top,\,\bot,\,a\neps b,\,a\subset b,\,a\notin b$ where \ $a,b$ are parameters in ${\cal M}$. We set~:

\smallskip\noindent
$\|\top\|=\vide$~; \ $\|\bot\|=\Pi$~; \ $\|a\neps b\|=\{\pi\in\Pi;\;(a,\pi)\in b\}$.

\smallskip\noindent
$\|a\subset b\|,\|a\notin b\|$ are defined simultaneously by induction on
$(\mbox{rk}(a)\cup\mbox{rk}(b),\mbox{rk}(a)\cap\mbox{rk}(b))$\\
($\mbox{rk}(a)$ being the rank of $a$ in ${\cal M}$).

\smallskip\noindent
$\dsp\|a\subset b\|=\bigcup_c\{\xi\ps\pi;\;\xi\in\Lbd,\;\pi\in\Pi,\;(c,\pi)\in a,\;
\xi\force c\notin b\}$~;

\smallskip\noindent
$\dsp\|a\notin b\|=\bigcup_c\{\xi\ps\xi'\ps\pi;\;\xi,\xi'\in\Lbd,\;\pi\in\Pi,\;(c,\pi)\in b,\;
\xi\force a\subset c,\;\xi'\force c\subset a\}$.

\smallskip\noindent
$\bullet$~~$F\equiv A\to B$~; then \
$\|F\|=\{\xi\ps\pi~;\;\xi\force A,\;\pi\in\|B\|\}$.

\smallskip\noindent
$\bullet$~~$F\equiv\pt x\,A$~: then \ $\dsp\|F\|=\bigcup_a\|A[a/x]\|$.

\smallskip\noindent
The following theorem, proved in~\cite{krivine6}, is an essential tool~:

\begin{theorem}[Adequacy lemma]\label{adequat}\ \\
Let $A_1,\ldots,A_n,A$ be closed formulas of \ZFe, and suppose that
$x_1:A_1,\ldots,x_n:A_n\vdash t:A$.\\
If \ $\xi_1\force A_1,\ldots,\xi_n\force A_n$ \ then \ $t[\xi_1/x_1,\ldots,\xi_n/x_n]\force A$.\\
In particular, if \ $\vdash t:A$, then \ $t\force A$.
\end{theorem}\noindent
Let $F$ be a closed formula of \ZFe, with parameters in ${\cal M}$. We say that
\emph{${\cal N}_{\cal A}$ realizes $F$} or that \emph{$F$ is realized in ${\cal N}_{\cal A}$}
(which is written \ ${\cal N}_{\cal A}\force F$ or even $\force F$), if there exists a proof-like term
$\theta$ such that \ $\theta\force F$.

\smallskip\noindent
It is shown in~\cite{krivine6} that \emph{all the axioms of \ZFe\ are
realized in ${\cal N}_{\cal A}$,} and thus also all the axioms of\, ZF\/.

\smallskip\noindent
{\bfseries Definitions.} Given a set of terms $X\subset\Lbd$ and a formula $F$, we shall use the notation
$X\to F$ as an \emph{extended formula}~; its truth value is \
$\|X\to F\|=\{\xi\ps\pi~;\;\xi\in X,\;\pi\in\|F\|\}$. 

\smallskip\noindent
Two formulas $F[x_1,\ldots,x_n]$ and $G[x_1,\ldots,x_n]$ of \ZFe\ will be called \emph{interchangeable}
if \ the formula $\pt x_1\ldots\pt x_n(F[x_1,\ldots,x_n]\dbfl G[x_1,\ldots,x_n])$ is realized.\\
\nopagebreak
That is, for instance, the case if \ $\|F[a_1,\ldots,a_n]\|=\|G[a_1,\ldots,a_n]\|$\\
\hspace*{8.6em}or also if \ $\|F[a_1,\ldots,a_n]\|=\|\neg\neg G[a_1,\ldots,a_n]\|$\\
for every $a_1,\ldots,a_n\in{\cal M}$.

\smallskip\noindent
The following lemma gives a useful example~:

\begin{lemma}\label{nAtoB}
For every formula $A$, define $^{\neg}A\subset\Lbd$ by \ \ $^{\neg}A=\{\kk_\pi\;;\;\pi\in\|A\|\}$.\\
Then \ $\neg A\to B$ and $^{\neg}A\to B$ are interchangeable, for every formula $B$.
\end{lemma}\noindent
We have immediately \ $\kk_\pi\force\neg A$ for every $\pi\in\|A\|$. Therefore,
$\|^{\neg}A\to B\|\subset\|\neg A\to B\|$ and it follows that \
$\III\force(\neg A\to B)\to(^{\neg}A\to B)$.\\
Conversely, let $\xi,\eta\in\Lbd,\,\xi\force ^{\neg}A\to B\,,\eta\force\neg B$ and let $\pi\in\|A\|$.\\
We have $\xi\kk_\pi\force B$, thus $(\eta)(\xi)\kk_\pi\force\bot$ and therefore
$(\eta)(\xi)\kk_\pi\star\pi\in\bbot$.\\
It follows that $\theta\star\xi\ps\eta\ps\pi\in\bbot$
with $\theta=\lbd x\lbd y(\ccc)\lbd k(y)(x)k$.\\
Finally, we have shown that $\theta\force(^{\neg}A\to B)\to(\neg B\to A)$, from which
the result follows.

\qed

\subsection*{Equality and type-like sets}\noindent
The formula \ $x=y$ \ is, by definition, \ $\pt z(x\neps z\to y\neps z)$ \emph{(Leibniz equality)}.

\smallskip\noindent
If $t,u$ are $\ell$-terms and $F$ is a formula of \ZFe, with parameters in ${\cal M}$,
we define the formula \ $t=u\hto F$. \ When it is closed, its truth value is~:\\
$\|t=u\hto F\|=\|\top\|=\vide$ \ if \ ${\cal M}\models t\ne u$~; 
$\|t=u\hto F\|=\|F\|$ \ if \ ${\cal M}\models t=u$.\\
The formula \ $t=u\hto\bot$ \ is written \ $t\ne u$.\\
The formula \ $t_1=u_1\hto(t_2=u_2\hto\cdots\hto(t_n=u_n\hto F)\cdots)$ \ is written~:\\
$t_1=u_1,t_2=u_2,\ldots,t_n=u_n\hto F$.

\smallskip\noindent
The formulas \ $t=u\to F$ \ and \ $t=u\hto F$ are interchangeable, as is shown in the~:

\begin{lemma}\ \\
i)~~$\CCC\,\III\,\III\,\force\pt x\pt y\left((x=y\to F)\to(x=y\hto F)\right)$~;\\
ii)~~$\CCC\,\III\,\force\pt x\pt y\left((x=y\hto F)\to(x=y\to F)\right)$.
\end{lemma}\noindent
i)~~Trivial.\\
ii)~~Let $a,b$ be individuals~; let $\xi\force a=b\hto F$, $\eta\force a=b$ and $\pi\in\|F\|$.\\
We show that $\eta\star\xi\ps\pi\in\bbot$.\\
Let $c=\{(b,\pi)\}$~; by hypothesis on $\eta$, we have $\eta\force a\neps c\to b\neps c$. Since $\pi\in\|b\neps c\|$,
it suffices to show that $\xi\force a\neps c$. This is clear if $a\ne b$, since $\|a\neps c\|=\vide$ in this case.\\
If $a=b$, then $\xi\force F$, by hypothesis on $\xi$, thus $\xi\star\pi\in\bbot$~; but $\|a\neps c\|=\{\pi\}$
in this case,  and therefore $\xi\force a\neps c$.

\qed

\smallskip\noindent
We set $\gimel X=X\fois\Pi$ for every individual $X$ of ${\cal M}$~; we define the quantifier
$\pt x^{\gimel X}$ as follows~:\\
\centerline{$\|\pt x^{\gimel X}F[x]\|=\bigcup_{a\in X}\|F[a]\|$.}

\noindent
Of course, we set \ $\ex x^{\gimel X}F[x]\equiv\neg\pt x^{\gimel X}\neg F[x]$.

\smallskip\noindent
The quantifier $\pt x^{\gimel X}$ has the intended meaning, which is that the formulas \
$\pt x^{\gimel X}F[x]$ \ and \ $\pt x(x\eps\gimel X\to F[x])$ \ are interchangeable.
This is shown by the~:

\begin{lemma}\ \\
$\CCC\,\III\,\force\pt x^{\gimel X}F[x]\to\pt x^{\gimel X}\neg\neg F[x]$~;\\
$\ccc\force\pt x^{\gimel X}\neg\neg F[x]\to\pt x^{\gimel X}F[x]$~;\\
$\|\pt x^{\gimel X}\neg\neg F[x]\|=\|\pt x(\neg F[x]\to x\neps\gimel X)\|$.
\end{lemma}\noindent
Immediate.

\qed

\smallskip\noindent
Each \emph{functional} $f:{\cal M}^n\to{\cal M}$, defined in ${\cal M}$ by a formula of ZF with
parameters, gives a function symbol, that we denote also by $f$, and which has the same interpretation
in the realizability model ${\cal N}_{\cal A}$.

\begin{proposition}\label{horn_equ}\ \\
Let $t,t_1,\ldots,t_n,u,u_1,\ldots,u_n$ be $\ell$-terms, built with variables $x_1,\ldots,x_k$
and functional symbols of ${\cal M}$.\\
If \ ${\cal M}\models\pt x_1\ldots\pt x_k(t_1=u_1,\ldots,t_k=u_k\to t=u)$, then~:\\
$\III\,\force\pt x_1\ldots\pt x_k(t_1=u_1,\ldots,t_k=u_k\hto t=u)$.\\
If \ ${\cal M}\models(\pt x_1\in X_1)\ldots(\pt x_k\in X_k)(t_1=u_1,\ldots,t_k=u_k\to t=u)$, then~:\\
$\III\,\force\pt x_1^{\gimel X_1}\ldots\pt x_k^{\gimel X_k}(t_1=u_1,\ldots,t_k=u_k\hto t=u)$.
\end{proposition}\noindent
Trivial.

\qed

\begin{proposition}
If $f:X_1\fois\cdots\fois X_n\to Y$ is a function in ${\cal M}$, its interpretation in ${\cal N}_{\cal A}$
is a function $f:\gimel X_1\fois\cdots\fois\gimel X_n\to\gimel Y$.
\end{proposition}\noindent
Indeed, let $f',f'':{\cal M}^n\to{\cal M}$ be any two functionals which are extensions of the function~$f$ to the
whole of ${\cal M}^n$. By proposition~\ref{horn_equ}(ii), we have~:\\
$\III\,\force\pt x_1^{\gimel X_1}\ldots\pt x_k^{\gimel X_k}(f'(x_1,\ldots,x_k)=f''(x_1,\ldots,x_k))$. 

\qed

\smallskip\noindent
An important example is the set $2=\{0,1\}$ equipped with the trivial boolean functions, written $\et,\ou,\neg$.
The extension to ${\cal N}_{\cal A}$ of these operations gives a structure of Boolean algebra
on $\gimel2$. It is called the \emph{characteristic Boolean algebra} of the model ${\cal N}_{\cal A}$.

\subsubsection*{Conservation of well-foundedness}\noindent
Theorem~\ref{bien_fonde} says that every well founded relation in the ground model ${\cal M}$, gives a
well foun\-ded relation in the realizability model ${\cal N}$.

\begin{theorem}\label{bien_fonde}
Let $f:{\cal M}^2\to2$ be a function defined in the ground model ${\cal M}$ such that ${f(x,y)=1}$ is a well founded relation on ${\cal M}$.
Then, for every formula $F[x]$ of \ \ZFe \ with parameters in ${\cal M}$~:\\
$\Y\force\pt y\left(\pt x(f(x,y)=1\hto F[x])\to F[y]\right)\to\pt y\,F[y]$\\
with \ $\Y=AA$ \ and \ $A=\lbd a\lbd f(f)(a)af$ (or $A=(\WWW)(\BBB)(\BBB\WWW)(\CCC)\BBB$).
\end{theorem}\noindent
Let us fix $b\in X$ and let \ $\xi\force\pt y\left(\pt x(f(x,y)=1\hto F[x])\to F[y]\right)$.\\
We show\/, by induction on $b$, following the well founded relation $f(x,y)=1$, that~:\\
$\Y\star\xi\ps\pi\in\bbot$ for every $\pi\in\|F[b]\|$.\\
Thus, suppose that \ $\pi\in\|F[b]\|$~; since \ $\Y\star\xi\ps\pi\succ\xi\star\Y\xi\ps\pi$,
we need to show that \ $\xi\star\Y\xi\ps\pi\in\bbot$.
By hypothesis, we have \ $\xi\force\pt x(f(x,b)=1\hto F[x])\to F[ b]$~;\\
Thus, it suffices to show that $\Y\xi\force f(a,b)=1\hto F[a]$ \ for every \ $a\in X$.\\
This is clear if $f(a,b)\ne1$, by definition of~~$\hto$.\\
If $f(a,b)=1$, we must show \ $\Y\xi\force F[a]$, \ i.e. \ $\Y\star\xi\ps\varpi\in\bbot$ for every \
$\varpi\in\|F[a]\|$.\\
But this follows from the induction hypothesis.

\qed

\smallskip\noindent
{\small{\bfseries Remarks.}\\
i)~~If the function $f$ is only defined on a set $X$ in the ground model ${\cal M}$, we can apply
theorem~\ref{bien_fonde} to the extension $f'$ of $f$ defined by $f'(x,y)=0$ if $(x,y)\notin X^2$.\\
This shows that, in the realizability model ${\cal N}$, the binay relation $f(x,y)=1$ is well founded
on~$\gimel X$.\\
ii)~~We can use theorem~\ref{bien_fonde} to show that the axiom of foundation of \ZFe\ is realized in
${\cal N}_{\cal A}$.\\
Indeed, let us define $f:{\cal M}^2\to2$ by setting $f(x,y)=1\Dbfl\ex z((x,z)\in y)$.
The binary relation $f(x,y)=1$ is obviously well founded in ${\cal M}$. Now, we have \
$\III\,\force\pt x\pt y(f(x,y)\ne1\to x\neps y)$ because $\pi\in\|x\neps y\|\To f(x,y)=1$.
Thus, the relation $x\eps y$ is stronger than the relation $f(x,y)=1$, which is well founded in~${\cal N}_{\cal A}$
by theorem~\ref{bien_fonde}.}

\subsection*{Integers}\noindent
Let $\phi,\alpha\in\Lbd$ and $n\in\NN$~; we define $(\phi)^n\alpha\in\Lbd$ by setting $(\phi)^0\alpha=\alpha$~;
$(\phi)^{n+1}\alpha=(\phi)(\phi)^n\alpha$.\\
For $n\in\NN$, we define \ $\ul{n}=(\sig)^n\ul{0}$ with \ $\ul{0}=\KKK\III$ and \
$\sig=(\BBB\WWW)(\BBB)\BBB$~;\\
$\ul{n}$ is ``the integer $n$'' and $\sig$ the ``successor'' in combinatory logic.\\
The essential property of $\ul{0}$ and $\sig$ is~: \
$\ul{0}\star\phi\ps\alpha\ps\pi\succ\alpha\star\pi$~; $\sig\star\nu\ps\phi\ps\alpha\ps\pi\succ\nu\star\phi\ps\phi\alpha\ps\pi$.

\smallskip\noindent
The following lemmas~\ref{phi^n-alpha} and \ref{phi^n-alpha-delta} will be used in section~\ref{gkappaden}.

\begin{lemma}\label{phi^n-alpha}\ \\
Let \ $O,\vsig\in\Lbd$ be such that~: \
$O\star\phi\ps\alpha\ps\pi\succc\alpha\star\pi$ \ and \
$\vsig\star\nu\ps\phi\ps\alpha\ps\pi\succc\nu\star\phi\ps\phi\alpha\ps\pi$\\
for every $\alpha,\zeta,\nu,\phi\in\Lbd$ and $\pi\in\Pi$.\\
Then, for every $n\in\NN,\alpha,\zeta,\phi\in\Lbd$ and $\pi\in\Pi$~:\\
i)~~$(\vsig)^nO\star\phi\ps\alpha\ps\pi\succc(\phi)^n\alpha\star\pi$~; in particular, \
$\ul{n}\star\phi\ps\alpha\ps\pi\succc(\phi)^n\alpha\star\pi$\\
ii)~~$(\vsig)^nO\star\CCC\BBB\phi\ps\zeta\ps\alpha\ps\pi\succc\zeta\star(\phi)^n\alpha\ps\pi$.
\end{lemma}\noindent
i)~~Proof by recurrence on $n$~; this is clear if $n=0$~; if $n=m+1$, we have~:\\
$\vsig\star(\vsig)^mO\ps\phi\ps\alpha\ps\pi\succc(\vsig)^mO\star\phi\ps\phi\alpha\ps\pi\succc
(\phi)^m(\phi)\alpha\star\pi$ by the recurrence hypothesis.\\
The particular case is $O=\ul{0},\vsig=\sig$.\\
ii)~~By (i), we have $(\vsig)^nO\star\CCC\BBB\phi\ps\zeta\ps\alpha\ps\pi\succc
(\CCC\BBB\phi)^n\zeta\star\alpha\ps\pi$.\\
We now show, by recurrence on $n$, that \ $(\CCC\BBB\phi)^n\zeta\star\alpha\ps\pi\succc
\zeta\star(\phi)^n\alpha\ps\pi$.\\
This is clear if $n=0$~; if $n=m+1$, we have~:\\
$(\CCC\BBB\phi)^n\zeta\star\alpha\ps\pi\succc\CCC\BBB\phi\star(\CCC\BBB\phi)^m\zeta\ps\alpha\ps\pi\succc
\CCC\star\BBB\ps\phi\ps(\CCC\BBB\phi)^m\zeta\ps\alpha\ps\pi\succc\\
\BBB\star(\CCC\BBB\phi)^m\zeta\ps\phi\ps\alpha\ps\pi\succc(\CCC\BBB\phi)^m\zeta\star\phi\alpha\ps\pi\succc
\zeta\star(\phi)^m(\phi)\alpha\ps\pi$ (by the recurrence hypothesis).

\qed

\begin{lemma}\label{phi^n-alpha-delta}\ \\
Let \ $\Omega,\Sigma\in\Lbd$ be such that~: \
$\Omega\star\delta\ps\phi\ps\alpha\ps\pi\succc\alpha\star\pi$ \ and \
$\Sigma\star\nu\ps\delta\ps\phi\ps\alpha\ps\pi\succc\nu\star\delta\ps\phi\ps\phi\alpha\ps\pi$\\
for every $\alpha,\delta,\nu,\phi\in\Lbd$ and $\pi\in\Pi$. For instance~:
$\Omega=(\KKK)(\KKK)\III$~; $\Sigma=(\BBB)(\BBB\WWW)(\BBB)\BBB$.\\
Then, for every $n\in\NN,\alpha,\delta,\zeta,\phi\in\Lbd$ and $\pi\in\Pi$~:\\
i)~~$(\Sigma)^n\Omega\star\delta\ps\phi\ps\alpha\ps\pi\succ(\phi)^n\alpha\star\pi$.\\
ii)~~$(\Sigma)^n\Omega\star\delta\ps\CCC\BBB\phi\ps\zeta\ps\alpha\ps\pi\succ\zeta\star(\phi)^n\alpha\ps\pi$.
\end{lemma}\noindent
Same proof as lemma~\ref{phi^n-alpha}.

\qed

\smallskip\noindent
We set \ $\NN_{\cal A}=\{(n,\ul{n}\ps\pi)\;;\;n\in\NN,\pi\in\Pi\}$~; it is shown below that
$\NN_{\cal A}$ is the set of integers of the realizability model ${\cal N}_{\cal A}$.

\smallskip\noindent
We define the quantifier \ $\pt x\indi$ as follows~:

\centerline{$\|\pt x\indi F[x]\|=\{\ul{n}\ps\pi\;;\;n\in\NN,\,\pi\in\|F[n]\|\}$.}

\noindent
that is also~:

\centerline{$\|\pt x\indi F[x]\|=\|\pt n^{\gimel\NN}(\{\ul{n}\}\to F[n])\|$.}

\smallskip\noindent
The formulas $\pt x\indi F[x]$ and $\pt x(x\eps\NN_{\cal A}\to F[x])$ are interchangeable, as is shown in the~:

\begin{lemma}\ \\
$\lbd x\lbd n\lbd y(y)(x)n\force\pt x\indi F[x]\to\pt x\indi\neg\neg F[x]$~;\\
$\lbd x\lbd n(\ccc)(x)n\force\pt x\indi\neg\neg F[x]\to\pt x\indi F[x]$~;\\
$\|\pt x\indi\neg\neg F[x]\|=\|\pt x(\neg F[x]\to x\neps\NN_{\cal A})\|$.
\end{lemma}\noindent
Immediate

\qed

\smallskip\noindent

\begin{lemma}\label{NA_ent}\ \\
i)~~$\KKK\force\pt x(x\neps\gimel\NN\to x\neps\NN_{\cal A})$.\\
ii)~~$\lbd x(x)\ul{0}\force 0\neps\NN_{\cal A}\to\bot$~; \
$\lbd f\lbd x(f)(\sig)x\force\pt y^{\gimel\NN}((y+1)\neps\NN_{\cal A}\to y\neps\NN_{\cal A})$.\\
iii)~~$\III\,\force\pt x\indi\left(\pt y^{\gimel\NN}(F[y]\to F[y+1]),F[0]\to F[x]\right)$ \
for every formula $F[x]$ of \ZFe.
\end{lemma}\noindent
i) and ii)~~Immediate.\\
iii)~~Let $n\in\NN,\,\phi\force\pt y^{\gimel\NN}(F[y]\to F[y+1]),\alpha\force F[0]$ et
$\pi\in\|F[n]\|$. We must show~:\\
$\ul{n}\star\phi\ps\alpha\ps\pi\in\bbot$ i.e., by lemma~\ref{phi^n-alpha},
$(\phi)^n\alpha\star\pi\in\bbot$.\\
But it is clear, by recurrence on $n$, that $(\phi)^n\alpha\force F[n]$ for every $n\in\NN$.

\qed

\smallskip\noindent
Lemma~\ref{NA_ent}(i) shows that $\NN_{\cal A}$ is a subset of $\gimel\NN$.\\
But it is clear that $\gimel\NN$ contains~$0$ and is closed by the function $n\mapsto n+1$.\\
Now, by lemma~\ref{NA_ent}(ii) and (iii), $\NN_{\cal A}$ is the smallest subset
of $\gimel\NN$ which contains $0$ and is closed by the function $n\mapsto n+1$. Therefore~:\\
\emph{$\NN_{\cal A}$ is the set of integers of the model ${\cal N}_{\cal A}$.}

\section{The characteristic Boolean algebra $\gimel2$}

\subsection*{Function symbols}\noindent
Let us now define the principal function symbols commonly used in the sequel~:

\smallskip\noindent
$\bullet$~~The projections \ $pr_0:X\fois Y\to X$ \ and \ $pr_1:X\fois Y\to Y$ \ defined by~:\\
$pr_0(x,y)=x,\;pr_1(x,y)=y$\\
give, in ${\cal N}_{\cal A}$, a \emph{bijection} from $\gimel(X\fois Y)$ onto $\gimel X\fois\gimel Y$.

\smallskip\noindent
$\bullet$~~We define, in ${\cal M}$, the function \ app $:Y^X\fois X\to Y$ (read \emph{application})
by setting~:\\
app$(f,x)=f(x)$ for $f\in Y^X$ and $x\in X$.\\
This gives, in ${\cal N}_{\cal A}$, an application \ app $:\gimel(Y^X)\fois\gimel X\to \gimel Y$.\\
We shall write $f(x)$ for \ app$(f,x)$.

\begin{theorem}\label{app_inj}\ \\
If \ $X\ne\vide$, the function \emph{app} gives an injection from $\gimel(Y^X)$ into \ $(\gimel Y)^{\gimel X}$.
Indeed, we have~:\\
$\III\,\force\pt f^{\gimel(Y^X)}\pt g^{\gimel(Y^X)}
\left(\pt x^{\gimel X}(\mbox{app}(f,x)=\mbox{app}(g,x))\to f=g\right)$.
\end{theorem}\noindent
Let $f,g\in Y^X$, $\xi\force\pt x^{\gimel X}(\mbox{app}(f,x)\ne\mbox{app}(g,x)\to\bot)$ and
$\pi\in\|f\ne g\to\bot\|$.\\
We must show $\xi\star\pi\in\bbot$. We choose $a\in X$~; then $\xi\force(f(a)\ne g(a)\to\bot)$.\\
If $f=g$, we have $\|f(a)\ne g(a)\to\bot\|=\|f\ne g\to\bot\|=\|\bot\to\bot\|$. Hence the result.\\
If $f\ne g$, we could choose $a$ such that $f(a)\ne g(a)$.\\
Then, $\|f(a)\ne g(a)\to\bot\|=\|f\ne g\to\bot\|=\|\top\to\bot\|$. Hence the result.

\qed

\smallskip\noindent
$\bullet$~~Let \ $\mbox{sp}:{\cal M}\to\{0,1\}$ (read \emph{support}) the unary function symbol  defined by~:\\
sp$(\vide)=0$~; sp$(x)=1$ if $x\ne\vide$.\\
In the realizability model ${\cal N}_{\cal A}$, we have  \ sp $:{\cal N}\to\gimel2$.

\smallskip\noindent
$\bullet$~~Let \ $\PP:\{0,1\}\fois{\cal M}\to{\cal M}$ (read \emph{projection}) the binary function symbol
defined by~:\\
$\PP(0,x)=\vide$~; $\PP(1,x)=x$.\\
In the realizability model ${\cal N}_{\cal A}$, we have  \
$\PP:\gimel2\fois{\cal N}\to{\cal N}$.\\
In the following, we shall write \ $ix$ instead of $\PP(i,x)$.\\
When $t,u$ are $\ell$-terms with values in $\gimel2$, we write $t\le u$ for $t\et u=t$.

\begin{proposition}\label{prop1_P}\ \\
i)~~$\III\,\force\pt i^{\gimel2}\pt x(i(jx)=(i\et j)x)$.\\
ii)~~$\III\,\force\pt i^{\gimel2}\pt x(ix=x\;\rightleftharpoons\mbox{ sp}(x)\le i)$.\\
iii)~~If \ $\vide\in E$, then \
$\III\,\force\pt i^{\gimel 2}\pt x^{\gimel E}(ix\eps\gimel E)$.\\
iv)~~If $f:{\cal M}^n\to{\cal M}$ is a function symbol  such that
$f(\vide,\ldots,\vide)=\vide$, then~:\\
\hspace*{1.8em}$\III\,\force\pt j^{\gimel2}\pt x_1\ldots\pt x_n
\left(jf(x_1,\ldots,x_n)=f(jx_1,\ldots,jx_n)\right)$.\\
v)~~$\III\,\force\pt i^{\gimel 2}\pt x\left(i\ne1\to\pt y(y\neps ix)\right)$ and therefore \
$\KKK^2\III\force\pt i^{\gimel 2}\pt x\left(i\ne1\to\pt y(y\notin ix)\right)$.
\end{proposition}\noindent
Trivial.

\qed

\smallskip\noindent
{\small{\bfseries Remark}. Proposition~\ref{prop1_P}(v) shows that, in the realizability model ${\cal N}$,
every non empty individual has support 1.}

\smallskip\noindent
Because of property (iv), we shall define, as far as possible, each function symbol
$f$ in~${\cal M}$, so that to have $f(\vide,\ldots,\vide)=\vide$.

\smallskip\noindent
$\bullet$~~Thus, let us change the ordered pair $(x,y)$ by setting $(\vide,\vide)=\vide$.
Then, we have~:\\
$\III\,\force\pt i^{\gimel2}\pt x\pt y\left(i(x,y)=(ix,iy)\right)$.

\smallskip\noindent
$\bullet$~~We define the binary function symbol \ $\sqcup:{\cal M}^2\to{\cal M}$ \ by setting~: \ $a\sqcup b=a\cup b$.

\smallskip\noindent
{\small{\bfseries Remark}. The extension to ${\cal N}$ of this operation \emph{is not} the union $\cup$.}

\smallskip\noindent
$\bullet$~~We define the \emph{strong inclusion symbol} $\subseteq$~: \ $x\subseteq y\equiv\pt z(z\neps y\to z\neps x)$.

\subsubsection*{The operation $\gimel_i$}\noindent
Let $E\in{\cal M}$ be such that $\vide\in E$. In ${\cal M}$, we define $\gimel_iE$ for $i\in2$
by setting~:\\
$\gimel_0E=\gimel\{\vide\}=\{\vide\}\fois\Pi$~; $\gimel_1E=\gimel E=E\fois\Pi$.\\
In this way, we have now defined $\gimel_iE$ in ${\cal N}$, for every $i\eps\gimel2$.

\begin{proposition}\label{prop2_P}\ \\
i)~~$\III\,\force\pt i^{\gimel2}\pt x\pt y(i(x\sqcup y)=ix\sqcup iy)$.\\
ii)~~$\III\,\force\pt i^{\gimel2}\pt j^{\gimel2}\pt x((i\ou j)x=ix\sqcup jx)$.\\
iii)~~$\III\,\force\pt i^{\gimel2}\pt j^{\gimel2}\pt x\pt y\pt z
(i\et j=0,z=ix\sqcup jy\hto iz=ix)$.\\
\hspace*{1.8em}$\III\,\force\pt i^{\gimel2}\pt j^{\gimel2}\pt x\pt y\pt z
(i\et j=0,z=ix\sqcup jy\hto jz=jy)$.\\
iv)~~$\III\,\force\pt i^{\gimel2}\pt j^{\gimel2}\pt x^{\gimel E}\pt y^{\gimel E}\pt z
\left(i\et j=0,z=ix\sqcup jy\hto z\eps\gimel_{i\ou j}E\right)$.
\end{proposition}\noindent
Trivial.

\qed

\begin{proposition}\label{propr_gimel_i}\ \\
If $\vide\in E,E'$, the following formulas are realized~:\\
i)~~$\gimel_iE$ increases with $i$. In particular, $\gimel_iE\subseteq\gimel E$.\\
ii)~~The $\veps$-elements of $\,\gimel_iE$ are the $ix$ for $x\eps\gimel E$.\\
iii)~~The $\veps$-elements of $\,\gimel_iE$ are those of $\gimel E$\  such that sp$(x)\le i$.\\
iv)~~The only $\veps$-element common to $\gimel_iE$ and $\gimel_{1-i}E$ is $\vide$.\\
v)~~If \ $i\et j=0$, then the application $x\mapsto(ix,jx)$ is a bijection from
$\gimel_{i\lor j}E$\\
\hspace*{1.5em}onto $\gimel_iE\fois\gimel_jE$.
The inverse function is \ $(x,y)\mapsto x\sqcup y$.\\
vi)~~$\gimel_i(E\fois E')=\gimel_iE\fois\gimel_iE'$.
\end{proposition}\noindent
We check immediately i), ii), iii), iv) below~:

\smallskip\noindent
i)~~$\III\,\force\pt i^{\gimel2}\pt j^{\gimel2}\pt x(i\et j=i\hto(x\neps\gimel_jE\to x\neps\gimel_iE))$.\\
ii)~~$\III\,\force\pt i^{\gimel2}\pt x^{\gimel E}(ix\eps\gimel_iE)$~;
$\III\,\force\pt i^{\gimel2}\pt x^{\gimel E}(ix\ne x\to x\neps\gimel_iE)$.\\
iii)~~$\III\,\force\pt i^{\gimel2}\pt x^{\gimel E}(x\neps\gimel_iE\to\mbox{sp}(x)i\ne\mbox{sp}(x))$~;
$\III\,\force\pt i^{\gimel2}\pt x^{\gimel E}(\mbox{sp}(x)i\ne\mbox{sp}(x)\to x\neps\gimel_iE)$~;\\
iv)~~$\III\,\force\pt i^{\gimel2}\pt x^{\gimel E}\pt y^{\gimel E}
(ix=(1-i)y\hto ix=\vide)$.

\smallskip\noindent
v)~~By proposition~\ref{prop2_P}(ii), we have \
$ix\sqcup jx=(i\lor j)x=x$ \ if \ $x\eps\gimel_{i\ou j}E$.\\
By proposition~\ref{prop2_P}(iii,iv), if $x,y\eps\gimel E$,
there exists \ $z\eps\gimel_{i\ou j}E$ \ such that $iz=ix,jz=jy$,
namely \ $z=ix\sqcup jy$.

\smallskip\noindent
vi)~~By proposition~\ref{prop1_P}(iv), we have \
$\III\,\force\pt i^{\gimel2}\pt x\pt y(i(x,y)=(ix,iy))$.

\qed

\begin{proposition}\label{EequipE'}
Let $E,E'\in{\cal M}$ be such that $\vide\in E,E'$ and $E$ is equipotent with $E'$. Then~:
$\force\pt i^{\gimel2}(\gimel_iE$ is equipotent with $\gimel_iE')$.
\end{proposition}\noindent
Let $\phi$ be, in ${\cal M}$, a bijection from $E$ onto $E'$, such that $\phi(\vide)=\vide$.
Then $\phi$ is, in ${\cal N}$, a bijection from $\gimel E$ onto $\gimel E'$. But we have
immediately~: $\III\,\force\pt i^{\gimel2}\pt x^{\gimel E}(\phi(ix)=i\phi(x))$.
This shows that $\phi$ is a bijection from $\gimel_iE$ onto $\gimel_iE'$.

\qed

\subsection*{Some general theorems}\noindent
Theorems~\ref{CF} to~\ref{k+ibf}, which are shown in this section, are valid \emph{in every realizability model}.

\smallskip\noindent
In the ground model ${\cal M}$, which satisfies \ ZFL, we denote by $\kappa$ the cardinal of $\Lbd\cup\Pi\cup\NN$
(which we shall also call the \emph{cardinal of the algebra ${\cal A}$}) and by $\kappa_+={\cal P}(\kappa)$
the power set of~$\kappa$.

\begin{theorem}\label{CF}\ \\
Let $\pt\vec{x}\pt y\,F[\vec{x},y]$ be a closed formula of \ZFe\ with parameters in ${\cal M}$
(where $\vec{x}=\!(x_1,\ldots,x_n)$).\\
Then, there exists in ${\cal M}$, a functional $f_F:\kappa\fois{\cal M}^n\to{\cal M}$ such that~:\\
i)~~If \ $\vec{a},b\in{\cal M}$ and $\xi\force F[\vec{a},b]$, then there exists $\alpha\in\kappa$
such that \ $\xi\force F[\vec{a},f_F(\alpha,\vec{a})]$.\\
ii)~~$\CCC\,\III\,\force\pt\vec{x}\pt y\left(F[\vec{x},y]\to
\ex\nu^{\gimel\kappa}F[\vec{x},f_F(\nu,\vec{x})]\right)$.
\end{theorem}\noindent
i)~~Let $\xi\mapsto\alpha_\xi$ be an injection from $\Lbd$ into $\kappa$.
Using the principle of choice in ${\cal M}$ (which satisfies $V=L$\,), we can define a functional \
$f_F:\kappa\fois {\cal M}^n\to{\cal M}$ \ such that, in ${\cal M}$, we have~: \
$\pt\vec{x}\pt y(\pt\xi\in\Lbd)
\left(\xi\force F[\vec{x},y]\Fl\xi\force F[\vec{x},f_F(\alpha_\xi,\vec{x})]\right)$.

\smallskip\noindent
ii)~~Let $\xi\force F[\vec{a},b]$, $\eta\force\pt \nu^{\gimel\kappa}\neg F[\vec{a},f_F(\nu,\vec{a})]$ and
$\pi\in\Pi$.\\
Thus, we have \ $\eta\force\neg F[\vec{a},f_F(\alpha_\xi,\vec{a})]$~;
by definition of $f_F$, we have  \ $\xi\force F[\vec{a},f_F(\alpha_\xi,\vec{a})]$. Therefore \ $\eta\star\xi\ps\pi\in\bbot$, and \
$\CCC\,\III\star\xi\ps\eta\ps\pi\in\bbot$.

\qed

\subsubsection*{Subsets of $\gimel\kappa_+$}

\begin{theorem}\label{Fsp(x)ne1}
Let $\pt x\pt y\pt z\,F[x,y,z]$ be a closed formula of \ZFe, with parameters in ${\cal M}$.\\
Then, there exists, in ${\cal M},$ a functional \ $\beta_F:{\cal M}\to\kappa_+$ such that~:\\
$\WWW\force\pt z\left(\pt x\pt y\pt y'(F[x,y,z],F[x,y',z]\to y=y')
\to\pt i^{\gimel2}\pt x(F[x,i\beta_F(z),z]\to\mbox{sp}(x)\ge i)\right)$.

\end{theorem}\noindent
By theorem~\ref{CF}(i), there exists, in ${\cal M}$, a functional \
$g:\kappa\fois{\cal M}^2\to{\cal M}$ such that~:\\
$(*)$~~For $a,b,c\in{\cal M}$ and $\xi\force F[a,b,c]$, there exists $\alpha\in\kappa$ such that \
$\xi\force F[a,g(\alpha,a,c),c]$.

\smallskip\noindent
Using the principle of choice in ${\cal M}$, we define a functional $\beta_F:{\cal M}\to\kappa_+$ such that~:\\
for every $\alpha\in\kappa$ and $c\in{\cal M}$, we have $\beta_F(c)\ne g(\alpha,\vide,c)$.\\
This is possible since $\kappa_+$ is of cardinal $>\kappa$.

\smallskip\noindent
Now let~: \ $a,c\in{\cal M}$, \ $i\in\{0,1\}$, \
$\phi\force\pt x\pt y\pt y'(F[x,y,c],F[x,y',c],y\ne y'\to\bot)$,\\
\hspace*{4.8em}$\xi\force F[a,i\beta_F(c),c]$, $\eta\force\mbox{sp}(a)i\ne i$ and $\pi\in\Pi$.

\smallskip\noindent
We must show that \ $\WWW\star\phi\ps\xi\ps\eta\ps\pi\in\bbot$, that is \
$\phi\star\xi\ps\xi\ps\eta\ps\pi\in\bbot$.\\
We set \ $b=i\beta_F(c)$ and therefore, we have \ $\xi\force F[a,b,c]$.\\
Thus, by $(*)$, we have \ $\xi\force F[a,g(\alpha,a,c),c]$
for some $\alpha\in\kappa$.

\smallskip\noindent
Let us show that $\|b\ne g(\alpha,a,c)\|\subset\|\mbox{sp}(a)i\ne i\|$~;
there are three possible  cases~:\\
If $i=0$, then $\|\mbox{sp}(a)i\ne i\|=\|0\ne0\|=\Pi$,  hence the result.\\
If $i=1$ and $a\ne\vide$, then $\|\mbox{sp}(a)i\ne i\|=\|1\ne1\|=\Pi$,  hence the result.\\
If $i=1$ and $a=\vide$, then~:\\
$\|b\ne g(\alpha,a,c)\|=\|i\beta_F(c)\ne g(\alpha,a,c)\|=\|\beta_F(c)\ne g(\alpha,\vide,c)\|
=\|\top\|=\vide$, by definition of~$\beta_F$,  hence the result.

\smallskip\noindent
It follows that \ $\eta\force b\ne g(\alpha,a,c)$. Now, we have  seen that~:\\
$\xi\force F[a,b,c]$ \ and \ $\xi\force F[a,g(\alpha,a,c),c]$.\\
Therefore, by hypothesis on $\phi$, we have \ $\phi\star\xi\ps\xi\ps\eta\ps\pi\in\bbot$.

\qed

\begin{corollary}\label{no_surj}
The following formulas are realized~:\\
i)~~$\pt E\pt i^{\gimel2}(\mbox{there is no surjection from }\bigcup\{\gimel_j E\;;\;j\eps\gimel2,j\not\geq i\}
\mbox{ onto }\gimel_i\kappa_+)$.\\
ii)~~$\pt E\pt i^{\gimel2}\pt j^{\gimel2}(\mbox{if there exists a surjection from }
\gimel_j E\mbox{ onto }\gimel_i\kappa_+\mbox{ then }j\ge i)$.\\
iii)~~$\pt i^{\gimel2}\pt j^{\gimel2}(i,j\ne0,i\et j=0\to$\\
\hspace*{\fill}$($there is no surjection from $\gimel_i\kappa_+\oplus\gimel_j\kappa_+
\mbox{ onto }\gimel_i\kappa_+\fois\gimel_j\kappa_+))$.
\end{corollary}\noindent
{\small{\bfseries Remark.}
The notation $\bigcup\{\gimel_j E\;;\;j\eps\gimel2,j\not\geq i\}$ denotes any individual $X$ of
${\cal N}$ such that~:\\
${\cal N}\models\pt x(x\eps X\dbfl\ex j^{\gimel2}(j\not\geq i\land x\eps\gimel_j E))$.}

\smallskip\noindent
i)~~We apply theorem~\ref{Fsp(x)ne1}, with the formula \ $F[x,y,z]\equiv(x,y)\eps z$.\\
In the realizabiblity model ${\cal N}$, we have  \ $\beta_F:{\cal N}\to\gimel\kappa_+$.\\
Let $z_0$ be, in ${\cal N}$, a surjective function onto \ $\gimel_i\kappa_+$.\\
We have \ $\beta_F(z_0)\eps\gimel\kappa_+$, and therefore \ $i\beta_F(z_0)\eps\gimel_i\kappa_+$.\\
If $x_0$ is such that $(x_0,i\beta_F(z_0))\eps z_0$, then \ sp$(x_0)\ge i$ \ by
theorem~\ref{Fsp(x)ne1}. Therefore, for any individual $E$, we have \ $x_0\eps\gimel_j E\;\Fl j\ge i$,
by proposition~\ref{propr_gimel_i}(iii).

\smallskip\noindent
ii)~~It is a trivial consequence of (i).

\smallskip\noindent
iii)~We take $E=\kappa_+$~; since $i,j\ne0,i\et j=0$, we have $i,j\not\ge i\ou j$~;
by (i), there is no surjection from $\gimel_i\kappa_+\cup\gimel_j\kappa_+$ onto
$\gimel_{i\ou j}\kappa_+$.\\
Now, since $i\et j=0$, $\gimel_{i\ou j}\kappa_+$ is equipotent with $\gimel_i\kappa_+\fois\gimel_j\kappa_+$
by proposition~\ref{propr_gimel_i}(v).\\
Moreover, $\vide$ is the only $\veps$-element common to $\gimel_i\kappa_+$ and $\gimel_j\kappa_+$ by proposition\ref{propr_gimel_i}(iv).\\
But these sets contain a countable subset by theorem~\ref{GEinfini}.
It follows that $\gimel_i\kappa_+\cup\gimel_j\kappa_+$ is equipotent with
$\gimel_i\kappa_+\oplus\gimel_j\kappa_+$.

\qed

\begin{theorem}\label{surj_2gk}
The formula~: $($there exists a surjection from $\gimel\kappa_+$ onto $2^{\gimel\kappa})$ \ is realized.
\end{theorem}\noindent
In the ground model ${\cal M}$, there exists a bijection from $\kappa_+=2^{\kappa}$ onto $2^{\kappa\fois\Pi}$.
Therefore, in~${\cal N}$, there exists a bijection from $\gimel\kappa_+$ onto $\gimel2^{\kappa\fois\Pi}$.\\
We now need a surjection from $\gimel2^{\kappa\fois\Pi}$ onto $2^{\gimel\kappa}$.\\
Let $\phi:{\cal M}\to2^{\kappa\fois\Pi}$ \ be the unary function symbol defined by \ $\phi(x)=x\cap(\kappa\fois\Pi)$.\\
In ${\cal N}$, we have  \ $\phi:{\cal N}\to\gimel2^{\kappa\fois\Pi}$. Now, we check immediately that~:

\smallskip\noindent
i)~~~$\III\,\force\pt\nu\pt x^{\gimel2^{\kappa\fois\Pi}}(\nu\neps\gimel\kappa\to\nu\neps x)$ \
because $\|\nu\neps a\|\subset\|\nu\neps\gimel\kappa\|$ for all $a\in{\cal P}(\kappa\fois\Pi)$.\\
ii)~~$\III\,\force\pt x\pt\nu^{\gimel\kappa}(\nu\neps x\rightleftharpoons\nu\neps\phi(x))$ \
because $\|\nu\neps a\|=\|\nu\neps\phi(a)\|$ for all $\nu\in\kappa$.

\smallskip\noindent
From (i), it follows that $\gimel2^{\kappa\fois\Pi}$ is, in ${\cal N}$, a set of subsets of $\gimel\kappa$~;\\
from (ii), it follows that  it contains at least one representative for each equivalence
class of extensionality.\\
Thus, the desired surjection simply associates, with each $\veps$-element of $\gimel2^{\kappa\fois\Pi}$,
its equivalence class of extensionality.

\qed

\begin{theorem}\label{GEinfini}
Let $E\in{\cal M}$ be infinite and such that $\vide\in E$. Then we have~:\\
$\force\pt i^{\gimel2}(i\ne0\to$ there exists an injection from $\NN$ into $\gimel_iE)$.
\end{theorem}\noindent
In ${\cal M}$, let \ $\phi:\NN\to(E\setminus\{\vide\})$ be injective. In ${\cal N}$, we have  \
$\phi:\gimel\NN\to\gimel E$.\\
The desired function is $n\mapsto i\phi(n)$. Indeed, we have~:\\
$\III\,\force\pt i^{\gimel2}\pt m^{\gimel\NN}\pt n^{\gimel\NN}
(i\ne0\to i\phi(m+n+1)\ne i\phi(m))$.\\
This shows that the restriction of this function to $\NN_{\cal A}$ (the set of integers of ${\cal N}_{\cal A}$)
is injective.

\qed

\begin{theorem}\label{kappa+nbo}
$\force\pt i^{\gimel2}\left(i\ne0,i\ne1\to(\gimel\kappa_+\mbox{ cannot be well ordered}\,)\right)$.
\end{theorem}\noindent
Let $i\in\gimel2,i\ne0,1$~; then, $\gimel_i\kappa_+$ and $\gimel_{1-i}\kappa_+$ are infinite
(theorem~\ref{GEinfini}) and $\subset\gimel\kappa_+$ by proposition~\ref{propr_gimel_i}(i).
But there exists no surjection from $\gimel_i\kappa_+$ onto $\gimel_{1-i}\kappa_+$, neither from
$\gimel_{1-i}\kappa_+$ onto $\gimel_i\kappa_+$, by corollary~\ref{no_surj}.

\qed

\smallskip\noindent
{\small{\bfseries Remark.} By theorem~\ref{kappa+nbo}, if the Boolean algebra $\gimel2$ is not trivial,
then $\gimel\kappa_+$ is not well orderable.
On the other hand, it can be shown that, if this Boolean algebra is trivial, then
the realizability model~${\cal N}$ is an extension by forcing of the ground model ${\cal M}$.
In this case, ${\cal N}$ itself can be well ordered, since we suppose that the ground model
${\cal M}$ satisfies ZFL.}

\subsubsection*{A strict order on $\gimel\kappa_+$}\noindent
A binary relation $<$ on $X$ is a \emph{strict order} if it is transitive ($x<y,y<z\Fl x<z$)
and antireflexive ($x\not<x$). This strict order is called \emph{total} if we have~:
$x<y$ or $y<x$ or $x=y$.

\smallskip\noindent
If $(X_0,<_0),(X_1,<_1)$ are two strictly ordered sets, then the \emph{strict order product} $<$ on
$X_0\fois X_1$ is defined by~: \ $(x_0,x_1)<(y_0,y_1)\Dbfl x_0<y_0$ and $x_1<y_1$.

\begin{lemma}\label{osbf}
The strict order product of $<_0,<_1$ is well founded if and only if one of the strict orders
$<_0,<_1$ is well founded.
\end{lemma}\noindent
Proof of $\Fl$~: by contradiction~; if $<_0$ and $<_1$ are not well founded, we have~:\\\nopagebreak
$\pt y_0\left(\pt x_0(x_0<_0y_0\to F_0[x_0])\to F_0[y_0]\right)$~; $\neg F_0[b_0]$~;\\
$\pt y_1\left(\pt x_1(x_1<_1y_1\to F_1[x_1])\to F_1[y_1]\right)$~; $\neg F_1[b_1]$~;\\
for some formulas $F_0,F_1$ and some individuals $b_0,b_1$. It follows that~:\\
$\pt y_0\pt y_1\left(\pt x_0\pt x_1(x_0<_0y_0,x_1<_1y_1\to G[x_0,x_1])\to G[y_0,y_1]\right)$~;
$\neg G[b_0,b_1]$\\
where $G[x_0,x_1]\equiv F_0[x_0]\lor F_1[x_1]$.

\smallskip\noindent
Proof of $\Leftarrow$~: suppose that $<_0$ is well founded and let $G[x_0,x_1]$ be any formula.\\
Let $F[x_0]\equiv\pt x_1G[x_0,x_1]$. We have to prove $\pt y_0\pt y_1G[y_0,y_1]$,
i.e. $\pt y_0F[y_0]$ with the hypothesis
$\pt y_0\pt y_1\left(\pt x_0\pt x_1(x_0<_0y_0,x_1<_1y_1\to G[x_0,x_1])\to G[y_0,y_1]\right)$.
But this implies~:\\
$\pt y_0\left(\pt x_0(x_0<_0y_0\to F[x_0])\to F[y_0]\right)$ and the result follows, because $<_0$
is well founded.

\qed

\smallskip\noindent
We denote by $\trgl$ a strict well ordering on $\kappa_+$, in ${\cal M}$~; we suppose that its least
element is~$\vide$ and that the cardinal of each proper initial segment is $\le\kappa$.

\smallskip\noindent
This gives a binary function from $\kappa_+^2$ into $\{0,1\}$, denoted by $(x\trgl y)$,
which is defined as follows~: $(x\trgl y)=1$ $\Dbfl$ $x\trgl y$.\\
We can extend it to the realizability model ${\cal N}_{\cal A}$, which gives a function from
$(\gimel\kappa_+)^2$ into~$\gimel2$.

\begin{lemma}\label{k+iprod}
The following propositions are realized~:\\
If $i\eps\gimel2,i\ne0$, then \ $(x\trgl y)=i$ is a strict ordering  of $\,\gimel_i\kappa_+$,
which we denote by $\trgl_i$.\\
If $i$ is an atom of the Boolean algebra $\gimel2$, then this ordering is total.
\end{lemma}\noindent
We have immediately~:\\
i)~~$\III\force\pt x^{\gimel\kappa_+}\pt y^{\gimel\kappa_+}\pt z^{\gimel\kappa_+}
((x\trgl y)\et(y\trgl z)\le(x\trgl z))$~; \
$\III\force\pt x^{\gimel\kappa_+}((x\trgl x)=0)$.\\
ii)~~$\III\,\force\pt i^{\gimel2}\pt x^{\gimel\kappa_+}\pt y^{\gimel\kappa_+}
\left((ix\trgl\,iy)\le i\right)$.\\
iii)~~$\III\,\force\pt x^{\gimel\kappa_+}\pt y^{\gimel\kappa_+}\left((x\trgl y)=0,(y\trgl x)=0\hto x=y\right)$.

\smallskip\noindent
It follows from (i) that, if $i\ne0$, then \ $(x\trgl y)\ge i$ is a strict ordering relation
on $\gimel\kappa_+$.\\
It follows from (ii), that this relation, restricted to $\gimel_i\kappa_+$, is equivalent to  \ $(x\trgl y)=i$.\\
Finally, it follows from (iii), that the relation \ $(x\trgl y)=i$, restricted to $\gimel_i\kappa_+$,
is total when $i$ is an atom of~$\gimel2$.

\qed

\begin{lemma}\label{k+ibf}
The following propositions are realized~:\\
i)~~$\pt i^{\gimel2}(\mbox{the application }x\mapsto(ix,(1-i)x)\,\mbox{is an isomorphism of strictly ordered sets}\\
\mbox{ from }(\gimel\kappa_+,\!\trgl)\mbox{ onto\ }(\gimel_i\kappa_+,\trgl_i)\fois(\gimel_{1-i}\kappa_+,\trgl_{1-i}))$.\\
ii)~~$\pt i^{\gimel2}($either $\gimel_i\kappa_+$ or $\gimel_{1-i}\kappa_+$ is a well founded ordered set$)$.
\end{lemma}\noindent
i)~It follows from proposition~\ref{propr_gimel_i}(v), that the application $x\mapsto(ix,(1-i)x)$ \
is a bijection from $\gimel\kappa_+$ onto ${\gimel_i\kappa_+\fois\gimel_{1-i}\kappa_+}$.
In fact, it is an \emph{isomorphism} of ordered sets, since we have~:

\smallskip\noindent
$\III\,\force\pt i^{\gimel2}\pt x^{\gimel\kappa_+}\pt y^{\gimel\kappa_+}
\left((x\trgl y)=(ix\trgl iy)\ou((1-i)x\trgl(1-i)y)\right)$ and therefore~:\\
$\force\pt i^{\gimel2}\pt x^{\gimel\kappa_+}\pt y^{\gimel\kappa_+}
\left((x\trgl y)=1\dbfl(ix\trgl iy)=i\land((1-i)x\trgl(1-i)y)=1-i\right)$.

\smallskip\noindent
ii)~By theorem~\ref{bien_fonde}, the relation $(x\trgl y)=1$ is well founded on $\gimel\kappa_+$.
Thus, the result follows immediately from (i) and lemma~\ref{osbf}.

\qed

\subsection*{$\gimel{\kappa}$ countable}\noindent
In this section, we consider some consequences of the hypothesis~: ($\gimel{\kappa}$ is countable).

\subsubsection*{Non extensional and dependent choice}\noindent
The formula $\pt z(z\neps y\to z\neps x)$ will be written $x\subseteq y$.\\
The formula $\pt x\pt y\pt y'((x,y)\eps f,(x,y')\eps f\to y=y')$ will be written Func$(f)$\\
(read~: \emph{$f$ is a function}).\\
We recall that $x\subseteq y$ is the formula $\pt z(z\neps y\to z\neps x)$.

\smallskip\noindent
The formula \
$\pt z\ex f\left(f\subseteq z\land\mbox{Func}(f)\land\pt x\pt y\ex y'((x,y)\eps z\to(x,y')\eps f)\right)$\\
is called the \emph{non extensional axiom of choice} and denoted by NEAC.

\smallskip\noindent
It is easily shown \cite{krivine6} that \  \ZFe + NEAC $\vdash\mbox{DC}$ (axiom of dependent choice).
On the other hand, we have  built, in \cite{krivine6}, a model of \ZFe\ + NEAC + $\neg$AC~;
and other such models will be given in the present paper. In all these models, $\RR$ is not well orderable.

\begin{theorem}\label{gimel_k_DC}\ \\
There exists a closed $\cc$-term \HHH\ such that \
$\HHH\force(\gimel\kappa\mbox{ is countable})\to\mbox{ NEAC}$.
\end{theorem}\noindent
We apply theorem~\ref{CF}(ii) to the formula \ $(x,y)\eps z$. We get a function symbol  $g$
such that \ $\CCC\,\III\force\pt x\pt y\pt z((x,y)\eps z\to\ex\nu^{\gimel\kappa}(x,g(\nu,x,z))\eps z)$.\\
Therefore, it suffices to prove NEAC in \ZFe, by means of this formula and the additional hypothesis~:
$(\gimel\kappa\mbox{ is countable})$. Now, from this hypothesis, it follows that there exists a strict
well ordering $<$ on $\gimel\kappa$. Then, we can define the desired function $f$ by means of the
comprehension  scheme~:\\
$(x,y)\eps f\dbfl(x,y)\eps z\land\ex\nu^{\gimel\kappa}\left(y=g(\nu,x,z)\land
\pt\alpha^{\gimel\kappa}(\alpha<\nu\to(x,g(\alpha,x,z))\neps z\right)$.\\
Intuitively, $f(x)=g(\nu,x,z)$ for the least $\nu\eps\gimel\kappa$ such that $(x,g(\nu,x,z))\eps z$.

\qed

\subsubsection*{Subsets of $\RR$}
\begin{theorem}\label{segdentrgl}
$\force(\gimel\kappa\mbox{ is countable})\,\to$\\
\hspace*{\fill}every bounded above subset of the ordered set $(\gimel\kappa_+,\trgl)$ is countable.
\end{theorem}\noindent

\smallskip\noindent
Every proper initial segment of the well ordering $\trgl$ on $\kappa_+$ is of cardinal $\kappa$. Thus,
there exists a function $\phi:\kappa\fois\kappa_+\to\kappa_+$ such that, for each $x\in\kappa_+,\,x\ne\vide$,
the function $\alpha\mapsto\phi(\alpha,x)$ is a surjection from $\kappa$ onto
$\{y\in\kappa_+\;;\;y\triangleleft x\}$. Then, we have immediately~:\\
$\III\,\force\pt x^{\gimel\kappa_+}\pt y^{\gimel\kappa_+}\left((y\trgl x)=1\hto
(\pt\alpha^{\gimel\kappa}(y\ne\phi(\alpha,x))\to\bot)\right)$.

\smallskip\noindent
This shows that, in ${\cal N}$, there exists a surjection from $\gimel\kappa$, onto every subset
of $\gimel\kappa_+$ which is bounded from above for the strict ordering $\trgl$.\\
Thus, all these subsets of $\gimel\kappa_+$ are countable, since $\gimel\kappa$ is.

\qed

\begin{theorem}\label{gk+R}
$\force(\gimel\kappa\mbox{ is countable}\,)\to$ there exists an injection from $\gimel\kappa_+$
into $\RR$.
\end{theorem}\noindent
We have obviously \ $\force(\gimel\kappa\mbox{ is countable}\,\to\gimel2\mbox{ is countable})$, and therefore~:\\
$\force(\gimel\kappa\mbox{ is countable}\,\to(\gimel2)^{\gimel\kappa}\mbox{ is equipotent to }\RR)$.\\
Now, by theorem~\ref{app_inj}, we have~: $\force(\mbox{there is an injection from }
\gimel\kappa_+=\gimel(2^\kappa)\mbox{ into }(\gimel2)^{\gimel\kappa})$.

\qed

\begin{theorem}\label{gk_den}
The following formula is realized~:\\
$(\gimel\kappa\mbox{ is countable}\,)\to$
there exists an application $i\mapsto X_i$ from the countable Boolean algebra $\gimel2$ into ${\cal P}(\RR)$
such that~:\\
i)~~$X_0=\{\vide\}$~; $i\ne0\,\to X_i$ is uncountably infinite~;\\
ii)~~$X_i\fois X_i$ is equipotent with $X_i$~;\\
iii)~~$X_i\cap X_j=X_{i\et j}$ \ and therefore \ $i\le j\,\to X_i\subset X_j$~;\\
iv)~~$i\et j=0\,\to X_{i\ou j}$ is equipotent with $X_i\fois X_j$~;\\
v)~~there exists a surjection from $X_1$ onto \ $\RR$.\\
vi)~~if $A$ is a subset of $\gimel2$ and if there is a
surjection from \ $\bigcup_{j\veps A}X_j$ onto $X_i$, then $i\le j$ for some $j\eps A$.\\
vii)~~if there is a surjection from $X_j$ onto $X_i$, then $i\le j$~;\\
viii)~~if $i,j\ne0,i\et j=0$, then there is no surjection from $X_i\oplus X_j$ onto $X_i\fois X_j$.
\end{theorem}\noindent
For each $i\eps\gimel2$, let us denote by  $X_i$ the image of $\gimel_i\kappa_+$ by
the injection from \ $\gimel\kappa_+$ into \ $\RR$, given by theorem~\ref{gk+R}.\\
i)~~The fact that $X_i$ is infinite for $i\ne0$ is a consequence of theorem~\ref{GEinfini}.\\
If $i=1$, $X_i$ is uncountable by (vi). If $i\ne0,1$ and $X_i$ is countable, then
$X_{1-i}$ is infinite and thus, there exists a surjection from $X_{1-i}$ onto $X_i$.
This contradicts corollary~\ref{no_surj}.\\
ii)~~by proposition~\ref{propr_gimel_i}(vi), $\gimel_i\kappa_+\fois\gimel_i\kappa_+$ is
equipotent with \ $\gimel_i(\kappa_+^2)$, thus also with \ $\gimel_i\kappa_+$ by proposition~\ref{EequipE'}.\\
iii)~~If $a\eps\gimel_i\kappa_+$ and $a\eps\gimel_j\kappa_+$, then $ia=a$, and therefore
$(i\et j)a=ja=a$.\\
iv)~~This is proposition~\ref{propr_gimel_i}(v).\\
v)~~Application of theorem~\ref{surj_2gk}.\\
vi), vii), viii)~~Applications of corollary~\ref{no_surj}.

\qed

\smallskip\noindent
Theorem~\ref{gk_den} is interesting only if the countable Boolean algebra $\gimel2$ is not trivial.
In this case, $\RR$ cannot be well ordered, by theorems~\ref{kappa+nbo} and~\ref{gk+R}.

\smallskip\noindent
In section~\ref{gkappaden} below, given an \emph{arbitrary} realizability algebra ${\cal A}$,
we build a new algebra ${\cal B}$ such that~:\\
$\bullet$~~${\cal N}_{\cal B}$ realizes the formula~: ($\gimel{\kappa}$ is countable).\\
$\bullet$~~The (countable) Boolean algebra $\gimel2$ of the model ${\cal N}_{\cal B}$ is elementarily
equivalent to the algebra $\gimel2$ of~${\cal N}_{\cal A}$.

\smallskip\noindent
In the sequel, we shall consider two interesting cases~:\\
$\gimel2$ is atomless~; $\gimel2$ has four $\veps$-elements.

\section{Collapsing $\gimel\kappa$}\label{gkappaden}
\subsection*{Extending a realizability algebra}
In the ground model ${\cal M}$, we consider a realizability algebra ${\cal A}$,
the elementary combinators of which are denoted by $\BBB,\CCC,\III,\KKK,\WWW,\ccc$
and the continuations \ $\kk_\pi$ for $\pi\in\Pi$.

\smallskip\noindent
We define the combinators \ $\BBB^*,\CCC^*,\III^*,\KKK^*,\WWW^*,\ccc^*,$ \
and the continuations \ $\kk^*_\pi$ as follows~: \

\smallskip\noindent
$\BBB^*=\lbd n\lbd x\lbd y\lbd z(xn)(\CCC)yz
=\bpptho(\CCC)(\BBB\CCC)(\CCC)(\BBB)(\BBB\BBB)\BBB\bppthf\CCC$~;\\
$\CCC^*=\lbd n\lbd x\lbd y\lbd z(x)nzy=(\CCC)(\BBB)\CCC$~;\\
$\III^*=\lbd n\lbd x(x)n=\CCC\,\III$~;\\
$\KKK^*=\lbd n\lbd x\lbd y(x)n=(\CCC)(\BBB)\KKK$~;\\
$\WWW^*=\lbd n\lbd x\lbd y(x)nyy=(\CCC)(\BBB)\WWW$~;\\
$\kk^*_\pi=\lbd n\lbd x(\kk_\pi)(x)n=(\CCC)(\BBB)\kk_\pi$~;\\
$\ccc^*=\lbd n\lbd x(\ccc)\lbd k(xn)\lbd n\lbd x(k)(x)n\\
=\bppptho(\CCC)\bppptho(\CCC)\bpptho(\BBB)\bptho(\BBB)(\BBB)\CCC\bpthf\CCC\bppthf(\CCC)(\BBB)
\bpptho(\BBB)(\BBB)\bptho(\BBB)(\BBB)\ccc\bpthf\BBB\bppthf\BBB\bpppthf\CCC\bpppthf\BBB$.

\smallskip\noindent
Therefore, we have~:

\smallskip\noindent
$\BBB^*\star\nu\ps\xi\ps\eta\ps\zeta\ps\pi\succc\xi\star\nu\ps\CCC\eta\zeta\ps\pi$~;\\
$\CCC^*\star\nu\ps\xi\ps\eta\ps\zeta\ps\pi\succc
\xi\star\nu\ps\zeta\ps\eta\ps\pi$~;\\
$\III^*\star\nu\ps\xi\ps\pi\succc\xi\star\nu\ps\pi$~;\\
$\KKK^*\star\nu\ps\xi\ps\eta\ps\pi\succc\xi\star\nu\ps\pi$~;\\
$\WWW^*\star\nu\ps\xi\ps\eta\ps\pi\succc\xi\star\nu\ps\eta\ps\eta\ps\pi$~;\\
$\kk^*_\pi\star\nu\ps\xi\ps\varpi\succc\xi\star\nu\ps\pi$~;\\
$\ccc^*\star\nu\ps\xi\ps\pi\succc\xi\star\nu\ps\kk^*_\pi\ps\pi$.

\smallskip\noindent
(reminder~: the notation \ $\xi\star\pi\succc\xi'\star\pi'$ \ means \
$\xi\star\pi\notin\bbot\Fl\xi'\star\pi'\notin\bbot$).

\smallskip\noindent
Let $\kappa$ be an infinite cardinal of ${\cal M}$, $\kappa\ge$ card$(\Lbd\cup\Pi)$~;
we consider the tree (usually called $\kappa^{<\omega}$) of functions, the domain of which is an integer, with values in $\kappa$.\\
Let $P$ be the ordered set obtained by adding a least element $\OO$ to this tree.\\
$P$ is an inf-semi-lattice, the greatest element $\1$ of which is the function $\vide$.\\
The greatest lower bound of $p,q\in P$, denoted by $pq$, is $p$ (resp. $q$) if $p,q\ne\OO$ and $q\subset p$
(resp. $p\subset q$). It is $\OO$ in every other case.

\smallskip\noindent
{\small{\bfseries Remark.} $P\setminus\{\OO\}=\kappa^{<\omega}$ is the ordered set used, in the method of forcing,
to \emph{collapse} (i.e. make countable) the cardinal $\kappa$.}

\smallskip\noindent
We define a new realizability algebra ${\cal B}$ by setting~:\\
$\LLbd=\Lbd\fois P\;;\;\PPi=\Pi\fois P\;;\;\LLbd\star\PPi=(\Lbd\star\Pi)\fois P$~;\\
$(\xi,p)\ps(\pi,q)=(\xi\ps\pi,pq)$~; \
$(\xi,p)\star(\pi,q)=(\xi\star\pi,pq)$~; \
$(\xi,p)(\eta,q)=(\CCC\xi\eta,pq)$.\\
{\bf B} $=(\BBB^*,\1)$~; {\bf C} $=(\CCC^*,\1)$~; {\bf I} $=(\III^*,\1)$~;
{\bf K} $=(\KKK^*,\1)$~; {\bf W} $=(\WWW^*,\1)$~;\\
$\cccc=(\ccc^*,\1)$~; \ $\kkk_{(\pi,p)}=(\kk^*_\pi,p)$.

\smallskip\noindent
We define, in ${\cal M}$, a function symbol from $P\fois\NN$ into $\{0,1\}$, denoted by
$(p\lle n)$, by setting~:\\
$(p\lle n)=1$ $\Dbfl$ $p\ne\OO$ and the domain of $p$ is an integer $\le n$.

\smallskip\noindent
We define $\bbot_{\cal B}$\,, that we shall denote also by $\bbbot$, as follows~:

\smallskip\noindent
$(\xi\star\pi,p)\in\bbbot$ $\Dbfl$ $(\pt n\in\NN)((p\lle n)=1\Fl\xi\star\ul{n}\ps\pi\in\bbot)$ \
for $p\in P\,,\,\xi\in\Lbd$ and $\pi\in\Pi$.\\
In particular, we have  $(\xi\star\pi,\OO)\in\bbbot$ for any \ $\xi\in\Lbd,\pi\in\Pi$.

\smallskip\noindent
We check now that ${\cal B}$ is a realizability algebra.

\smallskip\noindent
$\bullet$~~$(\xi,p)(\eta,q)\star(\pi,r)\notin\bbbot\Fl(\xi,p)\star(\eta,q)\ps(\pi,r)\notin\bbbot$~:\\
Suppose that \ $(\xi\star\eta\ps\pi,pqr)\in\bbbot$~; we must show \ $(\CCC\xi\eta\star\pi,pqr)\in\bbbot$ \
i.e. \ $\CCC\xi\eta\star\ul{n}\ps\pi\in\bbot$ for $(pqr\lle n)=1$. Now, we have  \
$\CCC\xi\eta\star\ul{n}\ps\pi\succc\xi\star\ul{n}\ps\eta\ps\pi$ \ which is in $\bbot$ by hypothesis.

\smallskip\noindent
$\bullet$~~$(\BBB^*,\1)\star(\xi,p)\ps(\eta,q)\ps(\zeta,r)\ps(\pi,s)\notin\bbbot$ $\Fl$
$(\xi,p)\star(\eta,q)(\zeta,r)\ps(\pi,s)\notin\bbbot$~:\\
Suppose that \ $(\xi,p)\star(\eta,q)(\zeta,r)\ps(\pi,s)\in\bbbot$ i.e.
$(\xi\star\CCC\eta\zeta\ps\pi,pqrs)\in\bbbot$.\\
We must show~:\\
$(\BBB^*\star\xi\ps\eta\ps\zeta\ps\pi,pqrs)\in\bbbot$ \ i.e. \
$\BBB^*\star\ul{n}\ps\xi\ps\eta\ps\zeta\ps\pi\in\bbot$ \ for $(pqrs\lle n)=1$.\\
Now, we have
$\BBB^*\star\ul{n}\ps\xi\ps\eta\ps\zeta\ps\pi\succc\xi\star\ul{n}\ps\CCC\eta\zeta\ps\pi$
which is in $\bbot$ by hypothesis.

\smallskip\noindent
$\bullet$~~$(\CCC^*,\1)\star(\xi,p)\ps(\eta,q)\ps(\zeta,r)\ps(\pi,s)\notin\bbbot$ $\Fl$
$(\xi,p)\star(\zeta,r)\ps(\eta,q)\ps(\pi,s)\notin\bbbot$~:\\
Suppose that \ $(\xi\star\zeta\ps\eta\ps\pi,pqrs)\in\bbbot$~; we must show~:\\
$(\CCC^*\star\xi\ps\eta\ps\zeta\ps\pi,pqrs)\in\bbbot$ \ i.e. \
$\CCC^*\star\ul{n}\ps\xi\ps\eta\ps\zeta\ps\pi\in\bbot$ \ for $(pqrs\lle n)=1$.\\
Now, we have \
$\CCC^*\star\ul{n}\ps\xi\ps\eta\ps\zeta\ps\pi\succc\xi\star\ul{n}\ps\zeta\ps\eta\ps\pi$ \
which is in $\bbot$ by hypothesis.

\smallskip\noindent
$\bullet$~~$(\III^*,\1)\star(\xi,p)\ps(\pi,q)\notin\bbbot$ $\Fl$
$(\xi,p)\star(\pi,q)\notin\bbbot$~:\\
Suppose that \ $(\xi\star\pi,pq)\in\bbbot$~; we must show~:\\
$(\III^*\star\xi\ps\pi,pq)\in\bbbot$ \ i.e. \
$\III^*\star\ul{n}\ps\xi\ps\pi\in\bbot$ \ for $(pq\lle n)=1$. Now, we have~:\\
$\III^*\star\ul{n}\ps\xi\ps\pi\succc\xi\star\ul{n}\ps\pi$ \
which is in $\bbot$ by hypothesis.

\smallskip\noindent
$\bullet$~~$(\KKK^*,\1)\star(\xi,p)\ps(\eta,q)\ps(\pi,r)\notin\bbbot$ $\Fl$
$(\xi,p)\star(\pi,r)\notin\bbbot$~:\\
Suppose that \ $(\xi\star\pi,pr)\in\bbbot$~; we must show~:\\
$(\KKK^*\star\xi\ps\eta\ps\pi,pqr)\in\bbbot$ \ i.e. \
$\KKK^*\star\ul{n}\ps\xi\ps\eta\ps\pi\in\bbot$ \ for $(pqr\lle n)=1$. Now, we have~:\\
$\KKK^*\star\ul{n}\ps\xi\ps\eta\ps\pi\succc\xi\star\ul{n}\ps\pi$ \
which is in $\bbot$ by hypothesis.

\smallskip\noindent
$\bullet$~~$(\WWW^*,\1)\star(\xi,p)\ps(\eta,q)\ps(\pi,r)\notin\bbbot$ $\Fl$
$(\xi,p)\star(\eta,q)\ps(\eta,q)\ps(\pi,r)\notin\bbbot$~:\\
Suppose that \ $(\xi\star\eta\ps\eta\ps\pi,pqr)\in\bbbot$~; we must show~:\\
$(\WWW^*\star\xi\ps\eta\ps\pi,pqr)\in\bbbot$ \ i.e. \
$\WWW^*\star\ul{n}\ps\xi\ps\eta\ps\pi\in\bbot$ \ for $(pqr\lle n)=1$. Now, we have~:\\
$\WWW^*\star\ul{n}\ps\xi\ps\eta\ps\pi\succc\xi\star\ul{n}\ps\eta\ps\eta\ps\pi$ \
which is in $\bbot$ by hypothesis.

\smallskip\noindent
$\bullet$~~$(\ccc^*,\1)\star(\xi,p)\ps(\pi,q)\notin\bbbot$ $\Fl$
$(\xi,p)\star(\kk^*_\pi,q)\ps(\pi,q)\notin\bbbot$~:\\
Suppose that \ $(\xi\star\kk^*_\pi\ps\pi,pq)\in\bbbot$~; we must show~:\\
$(\ccc^*\star\xi\ps\pi,pq)\in\bbbot$ \ i.e. \
$\ccc^*\star\ul{n}\ps\xi\ps\pi\in\bbot$ \ for $(pq\lle n)=1$.\\
Now, we have \ $\ccc^*\star\ul{n}\ps\xi\ps\pi\succc\xi\star\ul{n}\ps\kk^*_\pi\ps\pi$ \
which is in $\bbot$ by hypothesis.

\smallskip\noindent
$\bullet$~~$(\kk_\pi^*,p)\star(\xi,q)\ps(\varpi,r)\notin\bbbot$ $\Fl$
$(\xi,q)\star(\pi,p)\notin\bbbot$~:\\
Suppose that \ $(\xi\star\pi,pq)\in\bbbot$~; we must show~:\\
$(\kk^*_\pi\star\xi\ps\varpi,pqr)\in\bbbot$ \ i.e. \
$\kk^*_\pi\star\ul{n}\ps\xi\ps\varpi\in\bbot$ \ for $(pqr\lle n)=1$.\\
Now, we have \ $\kk^*_\pi\star\ul{n}\ps\xi\ps\varpi\succc\xi\star\ul{n}\ps\pi$ \
which is in $\bbot$ by hypothesis.

\smallskip\noindent
For each closed $\cc$-term $\tau$ (built with the elementary combinators and the application),
we define $\tau^*$ by recurrence, as follows~:\\
if $\tau$ is an elementary combinator, $\tau^*$ is already defined~;\\
we set \ $(tu)^*=\CCC t^*u^*$.

\smallskip\noindent
In the algebra ${\cal B}$, the value of the combinator $\tau$ is \
$\tau_{\cal B}=(\tau^*_{\cal A},\1)$.\\
In particular, the integer $n$ of the algebra ${\cal B}$ is \ $\ul{n}_{\cal B}=(\ul{n}^*,\1)$.\\
We have \ $\ul{0}_{\cal B}=(\ul{0}^*,\1)=(\KKK^*,\1)(\III^*,\1)$~; therefore~:
\hspace{9em}$\ul{0}^*=\CCC\KKK^*\III^*$.\\
We have \ $(\ul{n+1})_{\cal B}=((\ul{n+1})^*,\1)=(\sig^*,\1)(\ul{n}^*,\1)$~; therefore~:
\hspace{1em}$(\ul{n+1})^*=\C\sig^*\ul{n}^*$.

\smallskip\noindent
Thus, we have, for every $n\in\NN$~:\hspace{16.8em}$\ul{n}^*=(\C\sig^*)^n\ul{0}^*$.

\smallskip\noindent
We define the proof-like terms of the algebra ${\cal B}$ as terms of the form $(\theta,\1)$
where $\theta$ is a proof-like term of the algebra~${\cal A}$. The condition of coherence
for ${\cal B}$ is therefore~:\\
If \ $\theta$ is a proof-like term of ${\cal A}$, there exist $n\in\NN$ and $\pi\in\Pi$
such that $\theta\star\ul{n}\ps\pi\notin\bbot$.\\
\emph{If ${\cal A}$ is coherent, then so is ${\cal B}$}~: indeed, if $\theta$ is a proof-like term
of ${\cal A}$, then so is $\theta\ul{0}$~; this gives a stack $\pi$ such that $\theta\ul{0}\star\pi\notin\bbot$.

\smallskip\noindent
{\bfseries Notations.}\\
The realizability models associated with the algebras ${\cal A}$ and ${\cal B}$ are denoted
respectively by ${\cal N}_{\cal A}$ and~${\cal N}_{\cal B}$.\\
The truth value of a formula $F$ in the realizability model ${\cal N}_{\cal B}$ will be denoted by
$\|F\|_{\cal B}$ or also~$\vv F\vv$.\\
We write $(\xi,p)\force_{\cal B}F$ or $(\xi,p)\fforce F$ to say that $(\xi,p)$ realizes the  formula $F$ in the
realizability model~${\cal N}_{\cal B}$.

\subsection*{The collapsing function}
We now define ${\cal G}\in{\cal M}$ in the following way~:\\
${\cal G}=\{\left((m,\alpha),(\pi,p)\right);\;m\in\NN,\,\alpha\in\kappa,\,\pi\in\Pi,\,
p\in P\setminus\{\OO\},\,p(m)$ is defined and $p(m)=\alpha\}$.

\begin{theorem}\label{bij_ent_ind}\ \\
The formula \ $(\cal G$ is a surjection from \ $\NN$ onto \ $\gimel\kappa)$ is realized in
the model ${\cal N}_{\cal B}$.\\
More precisely, we have~:\\
i) $(\theta_0,\1)\fforce\pt x\pt y\pt y'
\left((x,y)\eps{\cal G},y\ne y'\to(x,y')\neps{\cal G}\right)$ \ with \ $\theta_0=\lbd n\lbd k\lbd x(x)n$~;\\
ii) $(\theta_1,\1)\fforce\pt y^{\gimel\kappa}[\pt x\indi((x,y)\neps{\cal G})\to\bot]$ \
with \ $\theta_1=\lbd n\lbd x((((n)(\CCC\BBB)(\CCC)\sig^*)(\CCC)x)\ul{0}^*)(\sig)n$,\\ 
and \ $\sig=(\BBB\WWW)(\BBB)\BBB$ (successor).
\end{theorem}\noindent
i) Let $m\in\NN,\alpha,\alpha'\in\kappa$, $(\pi,p)\in\vv(m,\alpha)\neps{\cal G}\vv$, \
$(\pi',p')\in\vv(m,\alpha')\neps{\cal G}\vv$\\
and $(\xi,q)\fforce\alpha\ne\alpha'$.\\
Thus, we have $m\in$ dom$(p),m\in$ dom$(p'),p(m)=\alpha$ and $p'(m)=\alpha'$.\\
By lemma~\ref{nAtoB}, we can replace the formula \ $(m,\alpha)\eps{\cal G}$, which is
$\neg((m,\alpha)\neps{\cal G})$, with the set of terms $^{\neg}((m,\alpha)\neps{\cal G})$ which is
$\{\kk_{(\pi,p)}\;;\;(\pi,p)\in\vv(m,\alpha)\neps{\cal G}\vv\}$.\\
Therefore, we have to show that~:\\
$(\theta_0,\1)\star\kk_{(\pi,p)}\ps(\xi,q)\ps(\pi',p')\in\bbbot$ \ that is \
$(\theta_0\star\kk^*_\pi\ps\xi\ps\pi',pp'q)\in\bbbot$.\\
This is obvious if $pp'q=\OO$. Otherwise, $p$ and $p'$ are compatible, thus \ $\alpha=\alpha'$.\\
Let $n$ be such that $(pp'q\lle n)=1$~; we must show that \ $\theta_0\star\ul{n}\ps\kk^*_\pi\ps\xi\ps\pi'\in\bbot$ \
i.e. \ $\xi\star\ul{n}\ps\pi'\in\bbot$.\\
Now, we have \ $(\xi,q)\fforce\bot$ by hypothesis on $(\xi,q)$, thus $(\xi,q)\star(\pi',\1)\in\bbbot$.\\
Since $(q\lle n)=1$, it follows that \ $\xi\star\ul{n}\ps\pi'\in\bbot$.

\smallskip\noindent
ii) Let us first show that \
$\theta_1\star\ul{n}\ps\eta\ps\varpi\succc\eta\star\ul{n+1}\ps\ul{n}^*\ps\varpi$ \
for each $n\in\NN,\eta\in\Lbd$ and $\varpi\in\Pi$.
We have \ $\theta_1\star\ul{n}\ps\eta\ps\varpi\succc
\ul{n}\star(\CCC\BBB)(\CCC)\sig^*\ps\CCC\eta\ps\ul{0}^*\ps\ul{n+1}\ps\varpi$.\\
By lemma~\ref{phi^n-alpha}(ii), in which we set $\zeta=\CCC\eta,\phi=\CCC\sig^*,\alpha=\ul{0}^*,\vsig=\sig,O=\ul{0}$ \
and \ $\pi=\ul{n+1}\ps\varpi$, we obtain~: \
$\theta_1\star\,\ul{n}\ps\eta\ps\varpi\succc\CCC\eta\star\ul{n}^*\ps\ul{n+1}\ps\varpi$ \
(since \ $\ul{n}^*=(\CCC\sig^*)^n\ul{0}^*$) \ $\succc\eta\,\star\,\ul{n+1}\ps\ul{n}^*\ps\varpi$.

\medskip\noindent
We prove now that \ $(\theta_1,\1)\fforce\pt y^{\gimel\kappa}[\pt x\indi((x,y)\neps{\cal G})\to\bot]$.\\
Let $\alpha\in\kappa$, $(\eta,p_0)\fforce\pt x\indi((x,\alpha)\neps{\cal G})$ and
$(\varpi,q_0)\in\Pi\fois P$~;\\
we show that \ $(\theta_1,\1)\star(\eta,p_0)\ps(\varpi,q_0)\in\bbbot$.\\
This is trivial if $p_0q_0=\OO$~;
otherwise, let $n\in\ennl$ be such that $(p_0q_0\lle n)=1$.\\
We must show that \ $\theta_1\star\ul{n}\ps\eta\ps\varpi\in\bbot$, that is \
$\eta\star\ul{n+1}\ps\ul{n}^*\ps\varpi\in\bbot$.\\
But we have  \ $(\eta,p_0)\fforce\{(\ul{n}^*,\1)\}\to(n,\alpha)\neps{\cal G}$ \
by hypothesis on $\eta$.\\
Since $(p_0q_0\lle n)=1$, we can define $q\in P$ with domain $n+1$ such that \ $q\supset p_0q_0$ \
and \ $q(n)=\alpha$.
Then, we have \ $(\varpi,q)\in\vv(n,\alpha)\neps{\cal G}\vv$ \  by definition of ${\cal G}$.\\
We have thus \ $(\eta,p_0)\star(\ul{n}^*,\1)\ps(\varpi,q)\in\bbbot$ that is \
$(\eta\star\ul{n}^*\ps\varpi,p_0q)\in\bbbot$.\\
But we have  $p_0q=q$, and therefore \  $(\eta\star\ul{n}^*\ps\varpi,q)\in\bbbot$.\\
Since \ $(q\lle n+1)=1$, it follows that \ $\eta\star\ul{n+1}\ps\ul{n}^*\ps\varpi\in\bbot$.

\qed

\begin{corollary}
${\cal N}_{\cal B}$ realizes the non extensional axiom of choice  and thus also DC.
\end{corollary}\noindent
Indeed, by theorem~\ref{bij_ent_ind}, the model ${\cal N}_{\cal B}$ realizes the formula~: \
($\gimel\kappa$ is countable).\\
But we have  $\kappa=$ card$(\LLbd\cup\PPi\cup\NN)$, since $\kappa\ge$  card$(\Lbd\cup\Pi\cup\NN)$
and $\kappa=$ card$(P)$.\\
Therefore \ ${\cal N}_{\cal B}$ realizes NEAC, by theorem~\ref{gimel_k_DC}.

\qed

\smallskip\noindent
{\small{\bfseries Remark.}
Intuitively, the model ${\cal N}_{\cal B}$ is an extension of the model
${\cal N}_{\cal A}$ obtained by forcing, by collapsing $\gimel\kappa$. We cannot apply
directly the usual theory of forcing, because $\gimel\kappa$ is not defined in~ZF\/.}

\subsection*{Elementary formulas}\noindent
Elementary formulas are defined as follows, where $t,u$ are $\ell$-terms, i.e. terms built with variables, individuals, and function symbols defined in ${\cal M}$:

\smallskip\noindent
$\bullet$~~$\top,\bot$ are elementary formulas~;\\
$\bullet$~~if $U$ is an elementary formula, then $t=u\hto U$ and $\pt x\,U$  are too~;\\
$\bullet$~~if $U,V$ are elementary formulas, then \ $U\to V$ too~;\\
$\bullet$~~if $U$ is an elementary formula, then $\pt n\indi U$ too.

\smallskip\noindent
{\small{\bfseries Remark.} $t\ne u$ is an elementary formula, and also $t\neps\gimel u$ (which can be written \
$f(t,u)\ne1$ where $f$ is the function symbol defined in ${\cal M}$ by~: \ $f(a,b)=1$ iff $a\in b$).\\
If $U$ is an elementary formula, then \ $\pt x^{\gimel t}U$ is too~: indeed, it is written \
$\pt x(f(x,t)=1\hto U)$.}

\smallskip\noindent
For each elementary formula $U$, we define two formulas $\,U_p$ and $U^p$,
with one additional free variable $p$, by the conditions below.\\
Condition~1 defines $U^p$ by means of $U_p$~; conditions 2 to 5 define $U_p$ by recurrence~:

\smallskip\noindent
1.~~$U^p\equiv\pt q^{\gimel P}\pt n\indi((pq\lle n)=1\hto U_q)$~;

\smallskip\noindent
2.~~$\bot_p\equiv\bot$ and $\top_p\equiv\top$~;

\smallskip\noindent
3.~~$(t=u\hto U)_p\equiv(t=u\hto U_p)$~; \
$(\pt x\,U[x])_p\equiv\pt x\,U_p[x]$~;

\smallskip\noindent
4.~~$(U\to V)_p\equiv\pt q^{\gimel P}\pt r^{\gimel P}
\left(p=qr\hto(U^q\to V_r)\right)$~;

\smallskip\noindent
5.~~$(\pt n\indi U[n])_p\equiv\pt n^{\gimel\NN}(\{\ul{n}^*\}\to U_p[n])$, in other words~:\\
\hspace*{1.4em}$\|(\pt n\indi U[n])_p\|=\{\ul{n}^*\ps\pi\;;\;n\in\NN,\pi\in\|U_p[n]\|\}$.

\begin{lemma}\label{(xi,p)fforceU}
For each closed elementary formula $U$, we have~:\\
$(\pi,p)\in\vv U\vv\Dbfl\pi\in\|U_p\|$~; \ $(\xi,p)\fforce U\Dbfl\xi\force U^p$.
\end{lemma}\noindent
Proof by recurrence on the length of the formula $U$.

\smallskip\noindent
1.~~We have \ $(\xi,p)\fforce U$ $\Dbfl$ $(\xi,p)\star(\pi,q)\in\bbbot$ for $(\pi,q)\in\vv U\vv$, that is~:\\
$(\xi\star\pi,pq)\in\bbbot$ for every $\pi\in\|U_q\|$, by the recurrence hypothesis, or also~:\\
$(\pt q\in P)(\pt\pi\in\|U_q\|)(\pt n\in\NN)((pq\lle n)=1\Fl\xi\star\ul{n}\ps\pi\in\bbot$) \
which is equivalent to~:\\
$\xi\force\pt q^{\gimel P}\pt n\indi((pq\lle n)=1\hto U_q)$ \ that is \
$\xi\force U^p$.

\smallskip\noindent
2 and 3.~~Obvious.

\smallskip\noindent
4.~~Any element of $\vv U\to V\vv$ has the form \ $(\xi,q)\ps(\pi,r)$, i.e. $(\xi\ps\pi,p)$, with
$p=qr$, $(\xi,q)\fforce U$ and $(\pi,r)\in\vv V\vv$~;\\
by the recurrence hypothesis, this is equivalent to \ $\xi\ps\pi\in\|U^q\to V_r\|$.

\smallskip\noindent
5.~~We have \ $\vv\pt n\indi U[n]\vv=\vv\pt n^{\gimel\NN}(\{(\ul{n}^*,\1)\}\to U[n])\vv$\\
$=\{(\ul{n}^*,\1)\ps(\pi,p)\;;\;n\in\NN,(\pi,p)\in\vv U[n]\vv\}
=\{(\ul{n}^*\!\ps\pi,p)\;;\;n\in\NN,(\pi,p)\in\vv U[n]\vv\}$.\\
Thus, by the recurrence hypothesis, it is \ $\{(\ul{n}^*\!\ps\pi,p)\;;\;n\in\NN,\pi\in\|U_p[n]\|\}$.

\qed

\begin{lemma}\label{theta_U^0}\ \\
For each elementary formula $U$, there exist two proof-like terms \ $\theta^0_U,\theta^1_U$,
such that~:\\
i)\hspace*{6.2em}$\theta^0_U\force\pt p^{\gimel P}\pt n\indi((p\lle n)=1\hto(U\to U_p))$~;\\
ii)\hspace*{5.9em}$\theta^1_U\force\pt p^{\gimel P}\pt n\indi((p\lle n)=1\hto(U_p\to U))$~;\\
iii)\hspace*{5.6em}$\tau^0_U\force\pt p^{\gimel P}\pt n\indi((p\lle n)=1\hto(U\to U^p))$~;\\
iv)\hspace*{5.7em}$\tau^1_U\force\pt p^{\gimel P}\pt n\indi((p\lle n)=1\hto(U^p\to U))$~;\\
with \ $\tau^0_U=\lbd n\lbd x\lbd m(\theta^0_U)mx$ \ and \ $\tau^1_U=\lbd n\lbd x(\theta^1_Un)(x)n$.
\end{lemma}\noindent
We first show (iii) and (iv) from (i) and (ii).

\smallskip\noindent
(i)$ \Fl $(iii)\\
Let $p\in P$ and $n\in\NN$ be such that $(p\lle n)=1$~; let $\xi\force U$ and $\pi\in\|U^p\|$.\\
We have to show~: \
$\lbd n\lbd x\lbd m(\theta^0_U)mx\star\ul{n}\ps\xi\ps\pi\in\bbot$.\\
Now, by the definition (1) of $U^p$, there exist $q\in P$, $m\in\NN$ and $\varpi\in\|U_q\|$ such that $(pq\lle m)=1$
and $\pi=\ul{m}\ps\varpi$. Therefore, we have $(q\lle m)=1$ and, by (i)~:\\
$\theta^0_U\star\ul{m}\ps\xi\ps\varpi\in\bbot$, hence \
$\lbd n\lbd x\lbd m(\theta^0_U)mx\star\ul{n}\ps\xi\ps\ul{m}\ps\varpi\in\bbot$.

\smallskip\noindent
(ii)$ \Fl $(iv)\\
Let $p\in P$, $n\in\NN,\xi\in\Lbd$ and $\pi\in\|U\|$ such that $(p\lle n)=1$ and $\xi\force U^p$.\\
We have to show~: \
$\lbd n\lbd x(\theta^1_Un)(x)n\star\ul{n}\ps\xi\ps\pi\in\bbot$ i.e.
$\theta^1_U\star\ul{n}\ps\xi\ul{n}\ps\pi\in\bbot$.\\
But, by the definition (1) of $U^p$, in which we set $q=p$, we have  $\xi\ul{n}\force U_p$~;
therefore, the desired result follows from (ii).

\smallskip\noindent
We now show (i) and (ii) by recurrence on the length of $U$.

\smallskip\noindent
$\bullet$~~If $U$ is $\bot$ or $\top$, we take $\theta^0_U=\theta^1_U=\lbd n\lbd x\,x$.

\smallskip\noindent
$\bullet$~~If $U\equiv(t=u\hto V)$ or $U\equiv\pt x\,V$, then $\theta^0_U=\theta^0_V$ and
$\theta^1_U=\theta^1_V$ by (3).

\smallskip\noindent
$\bullet$~~If $U\equiv V\to W$, let $q,r\in\NN$ and $p=qr$~; let $n\in\NN$ such that
$(p\lle n)=1$. We have~:\\
$\tau^0_V\ul{n}\force V\to V^q$~; \ \ $\tau^1_V\ul{n}\force V^q\to V$~; \ \
$\theta^0_W\ul{n}\force W\to W_r$~; \ \ $\theta^1_W\ul{n}\force W_r\to W$.

\smallskip\noindent
Let $\xi\force V\to W$~; then, by the recurrence hypothesis, we have~:\\
$(\theta^0_W\ul{n})\comp\xi\force V\to W_r$ \ and \
$(\theta^0_W\ul{n})\comp\xi\comp(\tau^1_V\ul{n})\force V^q\to W_r$.

\smallskip\noindent
Thus, by (4), we obtain \ $\theta^0_U=\lbd n\lbd x\lbd y(\theta^0_Wn)(x)(\tau^1_Vn)y$.

\smallskip\noindent
Now, let $\xi\force V^q\to W_r$~;  then, by the recurrence hypothesis, we have~:\\
$(\theta^1_W\ul{n})\comp\xi\force V^q\to W$ \ and \
$(\theta^1_W\ul{n})\comp\xi\comp(\tau^0_V\ul{n})\force V\to W$.

\smallskip\noindent
Thus, by (4), we obtain \ $\theta^1_U=\lbd n\lbd x\lbd y(\theta^1_Wn)(x)(\tau^0_Vn)y$.

\smallskip\noindent
$\bullet$~~If $U\equiv\pt n\indi V[n]$, we first prove~:

\begin{lemma}\label{n_n*}\ \\
There exist two proof-like terms $T_0,T_1$ such that, for every closed formula $F$ of \ZFe~:\\
i)~~$T_0\force\pt n^{\gimel\NN}((\{\ul{n}^*\}\to F)\to(\{\ul{n}\}\to F))$.\\
ii)~~$T_1\force\pt n^{\gimel\NN}((\{\ul{n}\}\to F)\to(\{\ul{n}^*\}\to F))$.\\
iii)~~For every elementary formula $V[n]$, we have~:\\
$T_0\force(\pt n\indi V[n])_p\to\pt n\indi V_p[n]$ \ and \
$T_1\force\pt n\indi V_p[n]\to(\pt n\indi V[n])_p$.
\end{lemma}\noindent
i)~~We apply lemma~\ref{phi^n-alpha}(ii) to the realizability algebra ${\cal A}$, with~:\\ $\vsig=\sig,\,O=\ul{0},\,\phi=\CCC\sig^*$ and $\,\alpha=\ul{0}^*$.
For every $n\in\NN,\zeta\in\Lbd$ and $\pi\in\Pi$, we obtain~:

\smallskip\noindent
$\ul{n}\star(\CCC\BBB)(\CCC)\sig^*\ps\zeta\ps\ul{0}^*\ps\pi\succc\zeta\star\ul{n}^*\ps\pi$, \
since \ $\ul{n}^*=(\C\sig^*)^n\ul{0}^*$.

\smallskip\noindent
Therefore, if we set \ $T_0=\lbd f\lbd n((n)(\CCC\BBB)(\CCC)\sig^*)f\ul{0}^*$, we have \
$T_0\star\zeta\ps\ul{n}\ps\pi\succc\zeta\star\ul{n}^*\ps\pi$.

\smallskip\noindent
Thus, we have \ $T_0\force\pt n^{\gimel\NN}((\{\ul{n}^*\}\to F)\to(\{\ul{n}\}\to F))$.

\smallskip\noindent
ii)~~We apply now lemma~\ref{phi^n-alpha}(i) to the realizability algebra ${\cal B}$, with~:\\
$\vsig=\sig_{\cal B},\,O=\ul{0}_{\cal B},\,\phi=(\CCC\Sigma,\1)\,\alpha=(\Omega,\1)$ and \
$\Omega=\lbd d\lbd f\lbd a\,a$~; $\Sigma=\lbd n\lbd d\lbd f\lbd a(ndf)(f)a$.

\smallskip\noindent
Since \  $\ul{n}_{\cal B}=(\sig_{\cal B})^n\ul{0}_{\cal B}=(\ul{n}^*,\1)$, we get,
by setting $\Sigma_2=(\CCC)^2\Sigma$~:\\
$(\ul{n}^*,\1)\star(\CCC\Sigma,\1)\ps(\Omega,\1)\ps(\varpi,\1)\succc
((\Sigma_2)^n\Omega,\1)\star(\varpi,\1)$\\
because $((\CCC\Sigma,\1))^n(\Omega,\1)=((\Sigma_2)^n\Omega,\1)$. We write this as~:\\
$(\ul{n}^*\star\CCC\Sigma\ps\Omega\ps\varpi,\1)\succc((\Sigma_2)^n\Omega\star\varpi,\1)$.\\
It follows that \ $\ul{n}^*\star\ul{0}\ps\CCC\Sigma\ps\Omega\ps\varpi\succc
(\Sigma_2)^n\Omega\star\ul{d}\ps\varpi$ \ for some $d\in\NN$.\\
Let us take $\varpi=\CCC\BBB\sig\ps\zeta\ps\ul{0}\ps\pi$. We obtain~:\\
$\ul{n}^*\star\ul{0}\ps\CCC\Sigma\ps\Omega\ps\CCC\BBB\sig\ps\zeta\ps\ul{0}\ps\pi\succc
(\Sigma_2)^n\Omega\star\ul{d}\ps\CCC\BBB\sig\ps\zeta\ps\ul{0}\ps\pi$.\\
Now, we apply lemma~\ref{phi^n-alpha-delta}(ii), with $\phi=\sig$ and $\alpha=\ul{0}$
(note that $\Sigma_2=(\CCC)^2\Sigma$ satisfies the hypothesis of lemma~\ref{phi^n-alpha-delta}).\\
We obtain \ $(\Sigma_2)^n\Omega\star\ul{d}\ps\CCC\BBB\sig\ps\zeta\ps\ul{0}\ps\pi\succ
\zeta\star(\sig)^n\ul{0}\ps\pi$ \ and therefore~:\\
$\ul{n}^*\star\ul{0}\ps\CCC\Sigma\ps\Omega\ps\CCC\BBB\sig\ps\zeta\ps\ul{0}\ps\pi\succc
\zeta\star\ul{n}\ps\pi$.\\
Finally, if we set \ $T_1=\lbd f\lbd n((((n\ul{0})(\CCC)\Sigma)\Omega)(\CCC)\BBB\sig)f\ul{0}$, we have~:\\
$T_1\star\zeta\ps\ul{n}^*\ps\pi\succc\zeta\star\ul{n}\ps\pi$ \ and therefore \
$T_1\force\pt n^{\gimel\NN}((\{\ul{n}\}\to F)\to(\{\ul{n}^*\}\to F))$.

\smallskip\noindent
iii)~~This follows immediately from (i) and (ii), by definition of \ $(\pt n\indi V[n])_p$.

\qed

\smallskip\noindent
We can now finish the proof of lemma~\ref{theta_U^0}, considering the last case which is~:\\
$\bullet$~~$U\equiv\pt m\indi V[m]$.

\smallskip\noindent
We show that $\theta^0_U=\lbd n\lbd x(T_1)\lbd m(\theta^0_Vn)(x)m$.\\
By the recurrence hypothesis, we have
$\theta^0_V\force\pt p^{\gimel P}\pt n\indi((p\lle n)=1\hto(V[m]\to V_p[m]))$.\\
Let $p\in P,n\in\NN,\xi\in\Lbd$ be such that $(p\lle n)=1$ and $\xi\force\pt m\indi V[m]$.\\
Then, for every $m\in\NN$, we have $\xi\ul{m}\force V[m]$~; thus $(\theta^0_V\ul{n})(\xi)m\force V_p[m]$
and therefore $\lbd m(\theta^0_V\ul{n})(\xi)m\force\pt m\indi V_p[m]$.
By lemma~\ref{n_n*}(iii), we get $(T_1)\lbd m(\theta^0_V\ul{n})(\xi)m\force(\pt m\indi V[m])_p$
and ~therefore~: $\lbd x(T_1)\lbd m(\theta^0_V\ul{n})(x)m\force\pt m\indi V[m]\to(\pt m\indi V[m])_p$. Finally~:\\
$\lbd n\lbd x(T_1)\lbd m(\theta^0_Vn)(x)m\force
\pt p^{\gimel P}\pt n\indi((p\lle n)=1\hto(\pt m\indi V[m]\to(\pt m\indi V[m])_p))$.

\smallskip\noindent
We show now that $\theta^1_U=\lbd n\lbd x\lbd m(\theta^1_Vn)(T_0)xm$.\\
By the recurrence hypothesis, we have
$\theta^1_V\force\pt p^{\gimel P}\pt n\indi((p\lle n)=1\hto(V_p[m]\to V[m]))$~;\\
Let $p\in P,n\in\NN,\xi\in\Lbd$ be such that $(p\lle n)=1$ and $\xi\force(\pt m\indi V[m])_p$.\\
By lemma~\ref{n_n*}(iii), we have $T_0\xi\force\pt m\indi V_p[m]$, thus $T_0\xi\ul{m}\force V_p[m]$.\\
Therefore $(\theta^1_V\ul{n})(T_0)\xi\ul{m}\force V[m]$, and
$\lbd m(\theta^1_V\ul{n})(T_0)\xi m\force\pt m\indi V[m]$, hence the result.

\qed

\begin{theorem}\label{elementaire}\ \\
The same closed elementary formulas, with parameters in ${\cal M}$, are realized in the models
${\cal N}_{\cal A}$ and ${\cal N}_{\cal B}$.
\end{theorem}\noindent
Let $U$ be a closed elementary formula, which is realized in ${\cal N}_{\cal A}$ and
let $\theta$ be a proof-like term such that \ $\theta\force U$.
Then, we have \ $(\tau^0_U)\ul{n}\theta\force U^p$ for $(p\lle n)=1$, by lemma~\ref{theta_U^0}(iii)~;\\
therefore, setting $p=\vide=\1$, we have  \ $((\tau^0_U)\ul{0}\theta,\1)\fforce U$
by lemma~\ref{(xi,p)fforceU}.\\
Therefore, the formula $U$ is  also realized in the model ${\cal N}_{\cal B}$.\\
Conversely, if $(\theta,\1)\fforce U$ with $\theta\in$ QP\/, we have \ $\theta\force U^{\1}$,
by lemma~\ref{(xi,p)fforceU}. Thus \
$\tau^1_U\ul{0}\,\theta\force U$ by lemma~\ref{theta_U^0}(iv).

\qed

\smallskip\noindent
{\small{\bfseries Remark.} For instance~:\\
$\bullet$~~If the Boolean algebra $\gimel2$ has four $\veps$-elements or if it is atomless,
in the model~${\cal N}_{\cal A}$, the same goes for the model ${\cal N}_{\cal B}$.\\
$\bullet$~~Arithmetical formulas are elementary. Therefore, by theorem~\ref{elementaire},
the models ${\cal N}_{\cal A}$ and ${\cal N}_{\cal B}$ realize the same  arithmetical formulas.
In fact, this was already known, because they are the same as the arithmetical formulas which are true
in ${\cal M}$~\cite{krivine3,krivine4}.
}

\subsection*{Arithmetical formulas  and dependent choice}\noindent
In this section, we obtain, by means of the previous results, a technique to transform into a
program, a given proof, in \ ZF + DC, of an arithmetical formula~$F$.\\
We notice that this program is a closed $\cc$-term, written with the elementary combinators
$\BBB,\CCC,\III,\KKK,\WWW,\ccc$ \emph{without any other instruction}. 

\smallskip\noindent
Thus, let us consider a proof of \ \ZFe\ $\vdash\mbox{ NEAC }\to F$. It gives us a closed $\cc$-term
$\Phi_0$ such that \ $\Phi_0\force\mbox{NEAC }\to F$, in every realizability algebra.\\
We now describe a rewriting on closed $\cc$-terms, which will transform $\Phi_0$ into
a closed $\cc$-term~$\Phi$ \ such that \ $\Phi\force F$ \ in every realizability algebra ${\cal A}$.

\smallskip\noindent
By theorem~\ref{gimel_k_DC}, we have  $\Phi_1\force(\gimel\kappa\mbox{ is countable})\to F$ \
with \ $\Phi_1=\lbd x(\Phi_0)(\HHH)x$.\\
We apply this result in the algebra ${\cal B}$, which gives~:\\
$(\Phi_1^*,\1)\fforce(\gimel\kappa\mbox{ is countable})\to F$.\\
Now, theorem~\ref{bij_ent_ind} gives a closed $\cc$-term $\Delta$ such that
$(\Delta,\1)\fforce(\gimel\kappa\mbox{ is countable})$.\\
It follows that $(\Phi_1^*,\1)(\Delta,\1)\fforce F$, i.e. $(\Psi,\1)\fforce F$,
with $\Psi=\CCC\Phi_1^*\Delta$.\\
Since $F$ is an arithmetical formula, it is an elementary formula.\\
Therefore, by lemma~\ref{(xi,p)fforceU}, we have \ $\Psi\force F^{\1}$.
Now, by lemma~\ref{theta_U^0}(iv), we have~:\\
$\tau^1_F\force\pt p^{\gimel P}\pt n\indi((p\lle n)=1\hto(F^p\to F))$.\\
We set $p=\1$ and $n=0$, and we obtain \ $\tau^1_F\ul{0}\force F^{\1}\to F$.

\smallskip\noindent
Finally, by setting \ $\Phi=(\tau^1_F)\ul{0}\Psi$, we have  \ $\Phi\force F$.

\subsection*{A relative consistency result}\noindent
In \cite{krivine6}, we have  defined a countable realizability algebra ${\cal A}$ such that
the characteristic Boolean algebra $\gimel2$ of the model ${\cal N}_{\cal A}$ is atomless
(in this example, we have $\kappa=\NN$).\\
If we apply the technique of section~\ref{gkappaden}, in order to collapse $\gimel\kappa$,
we obtain a realizability algebra ${\cal B}$ and a model ${\cal N}_{\cal B}$,
the characteristic Boolean algebra of which is also atomless. Indeed, the property~:
$(\gimel2\mbox{ is atomless})$ is expressed by an elementary formula.\\
But now $\gimel2$ is \emph{the countable atomless Boolean algebra} (they are all isomorphic).
Therefore, by applying theorems~\ref{gimel_k_DC} and~\ref{gk_den}, we obtain the relative consistency
result (i) announced in the introduction.

\smallskip\noindent
{\small{\bfseries Remark.} We note that this method applies to every realizability algebra such that
we have~:\\
$\force(\gimel2$ is an atomless Boolean algebra).}

\section{A two threads model ($\gimel2$ with four elements)}\noindent
In this section, we suppose that ${\cal A}$ is a \emph{standard realisability algebra}~\cite{krivine6}.\\
This means, by definition, that the terms and the stacks are finite sequences, built with~:

\smallskip\noindent
the alphabet $\BBB$, $\CCC$, $\III$, $\KKK$, $\WWW$, $\ccc$, $\kk$, $\ps$, $($, $)$, $[$, $]$\\
a countable set of \emph{term constants} (also called \emph{instructions}),\\
a countable set of \emph{stack constants}

\smallskip\noindent
and that they are defined by the following rules~:

\smallskip\noindent
 $\BBB$, $\CCC$, $\III$, $\KKK$, $\WWW$, $\ccc$ and all the term constants are terms~;\\
if $t,u$ are terms, the sequence $(t)u$ is a term~;\\
if $\pi$ is a stack, the sequence $\kk[\pi]$ is a term (denoted by $\kk_\pi$)~;\\
each stack constant is a stack~;\\
if $t$ is a term and $\pi$ is a stack, then $t\ps\pi$ is a stack.

\smallskip\noindent
If $t$ is a term and $\pi$ is a stack, then the ordered pair $(t,\pi)$ is a \emph{process},
denoted by \ $t\star\pi$.

\smallskip\noindent
A proof-like term of ${\cal A}$ is a term which does not contain the symbol $\kk$~; or, which is
the same, a term which does not contain any stack constant.

\smallskip\noindent
We now build a realizability model in which $\gimel2$ has exactly 4 elements.

\smallskip\noindent
We suppose that there are exactly two stack constants $\pi^0$, $\pi^1$  and one term constant $\dd$.\\
For $i\in\{0,1\}$, let $\Lbd^i$ (resp. $\Pi^i$) be the set of terms (resp. stacks)\\
which contain the only stack constant $\pi^i$.\\
For $i,j\in\{0,1\}$, define $\bbot^i_j$ as the least set $P\subset\Lbd^i\star\Pi^i$ of processes such that~:\\
1.~~$\dd\star\ul{j}\ps\pi\in P$ \ for every $\pi\in\Pi^i$.

\smallskip\noindent
2.~~$\xi\star\pi\in\Lbd^i\star\Pi^i$, $\xi'\star\pi'\in P$, $\xi\star\pi\succ\xi'\star\pi'\,$ $\Fl$ $\xi\star\pi\in P$

\smallskip\noindent
3.~~If at least two out of three processes $\xi\star\pi,\,\eta\star\pi,\,\zeta\star\pi$ are in $P$,
then $\dd\star\ul{2}\ps\xi\ps\eta\ps\zeta\ps\pi\in P$.

\smallskip\noindent
{\small{\bfseries Remarks.}\\
The preorder $\succ$ on $\Lbd\star\Pi$ was defined at the beginning of section~\ref{general}.\\
We express condition~2 by saying that \emph{$P$ is saturated in $\Lbd^i\star\Pi^i$}.\\
Following this definition of \ $\succ$, the constant $\dd$ is a \emph{halting instruction}.
Indeed, we have~:\\
\centerline{$\dd\star\pi\succ\xi\star\varpi$ \ $\Dbfl$ \ $\xi\star\varpi=\dd\star\pi$.}}

\medskip\noindent
We define $\bbot$ \ by~:\hspace{2em}{
$\Lbd\star\Pi\setminus\bbot=(\Lbd^0\star\Pi^0\setminus\bbot^0_0)\cup(\Lbd^1\star\Pi^1\setminus\bbot^1_1)$}\\
In other words, a process is in $\bbot$ if and only if\\
either it is in $\bbot^0_0\cup\bbot^1_1$ or it contains both stack constants $\pi^0,\pi^1$.

\begin{lemma}\label{bbot_ij_red}
If \ $\xi\star\pi\in\bbot^i_j$ and \ $\xi\star\pi\succ\xi'\star\pi'$ then \
$\xi'\star\pi'\in\bbot^i_j$ (closure by reduction).
\end{lemma}\noindent
Suppose that \ $\xi_0\star\pi_0\succ\xi'_0\star\pi'_0$~, \ $\xi_0\star\pi_0\in\bbot^i_j$ and \
$\xi'_0\star\pi'_0\notin\bbot^i_j$. We may suppose that~:

\smallskip\noindent
$(*)$\hspace{8em}$\xi_0\star\pi_0\succ\xi'_0\star\pi'_0$ in exactly one step of reduction.

\smallskip\noindent
Let us show that $\bbot^i_j\setminus\{\xi_0\star\pi_0\}$ has properties 1,2,3 defining $\bbot_j^i$\,, \
which will contradict the definition of $\bbot_j^i$~:\\
1.~~If $\xi_0\star\pi_0=\dd\star\ul{j}\ps\pi$, with $\pi\in\Pi^i$, \ then \ $\dd\star\ul{j}\ps\pi\succ\xi'_0\star\pi'_0$, \ thus \
$\xi'_0\star\pi'_0=\dd\star\ul{j}\ps\pi$.\\
Therefore \ $\xi'_0\star\pi'_0\in\bbot^i_j$\,, which is false.

\smallskip\noindent
2.~~Suppose \ $\xi\star\pi\in\Lbd^i\star\Pi^i$, $\xi\star\pi\succ\xi'\star\pi'\in\bbot^i_j$\,, $\xi'\star\pi'\ne\xi_0\star\pi_0$. Then \ $\xi\star\pi\in\bbot^i_j$, by~(2).\\
If \ $\xi\star\pi=\xi_0\star\pi_0$, then \ $\xi_0\star\pi_0\succ\xi'\star\pi'$~; since
$\xi'\star\pi'\ne\xi_0\star\pi_0$, it follows from $(*)$ that $\xi'_0\star\pi'_0\succ\xi'\star\pi'$
and therefore $\xi'_0\star\pi'_0\in\bbot^i_j$\,, which is false.

\smallskip\noindent
3.~~Suppose that two out of the processes $\xi\star\pi,\,\eta\star\pi,\,\zeta\star\pi$ are in
$\bbot^i_j\setminus\{\xi_0\star\pi_0\}$, but $\dd\star\ul{2}\ps\xi\ps\eta\ps\zeta\ps\pi$ is not.
From (3), it follows that \ $\dd\star\ul{2}\ps\xi\ps\eta\ps\zeta\ps\pi=\xi_0\star\pi_0$.\\
Thus,  \ $\dd\star\ul{2}\ps\xi\ps\eta\ps\zeta\ps\pi\succ\xi'_0\star\pi'_0$, and therefore \
$\xi'_0\star\pi'_0=\dd\star\ul{2}\ps\xi\ps\eta\ps\zeta\ps\pi$.

\smallskip\noindent
Therefore $\xi'_0\star\pi'_0\in\bbot^i_j$\,, which is false.

\smallskip
\qed

\begin{lemma}\label{bbot^i_0_1}
$\bbot^i_0\cap\bbot^i_1=\vide$.
\end{lemma}\noindent
We prove that \ $(\Lbd^i\star\Pi^i\setminus\bbot^i_1)\supset\bbot^i_0$ by showing that \
$\Lbd^i\star\Pi^i\setminus\bbot^i_1$ \ has properties 1, 2, 3 which define $\bbot^i_0$.

\smallskip\noindent
1. $\dd\star\ul{0}\ps\pi\notin\bbot^i_1$ because $\bbot^i_1\setminus\{\dd\star\ul{0}\ps\pi\}$
has properties 1, 2, 3 defining $\bbot^i_1$.\\
2. Follows from lemma~\ref{bbot_ij_red}.\\
3. Suppose $\xi_0\star\pi_0,\,\eta_0\star\pi_0\notin\bbot_1^i$~; we show that
$\dd\star\ul{2}\ps\xi_0\ps\eta_0\ps\zeta_0\ps\pi_0\notin\bbot^i_1$ \ by showing\\
that $\bbot^i_1\setminus\{\dd\star\ul{2}\ps\xi_0\ps\eta_0\ps\zeta_0\ps\pi_0\}$
has properties 1, 2, 3 defining $\bbot^i_1$.

\smallskip\noindent
\hspace*{1em}1.~~Clearly, \ $\dd\star\ul{1}\ps\pi'\in
(\bbot^i_1\setminus\{\dd\star\ul{2}\ps\xi_0\ps\eta_0\ps\zeta_0\ps\pi_0\})$ for every $\pi'\in\Pi^i$.

\smallskip\noindent
\hspace*{1em}2.~~Suppose that \ $\xi\star\pi\in\Lbd^i\star\Pi^i$, $\xi\star\pi\succ\xi'\star\pi'\in\bbot^i_1$,
$\xi'\star\pi'\ne\dd\star\ul{2}\ps\xi_0\ps\eta_0\ps\zeta_0\ps\pi_0$ and\\
\hspace*{2.4em}that $\xi\star\pi\notin(\bbot^i_1\setminus\{\dd\star\ul{2}\ps\xi_0\ps\eta_0\ps\zeta_0\ps\pi_0\})$.\\
\hspace*{2.4em}From (2), it follows that $\xi\star\pi=\dd\star\ul{2}\ps\xi_0\ps\eta_0\ps\zeta_0\ps\pi_0$
which contradicts $\xi\star\pi\succ\xi'\star\pi'$.

\smallskip\noindent
\hspace*{1em}3.~~Suppose that two out of the processes \ $\xi\star\pi,\,\eta\star\pi,\,
\zeta\star\pi$ are in $\bbot^i_1\setminus\{\dd\star\ul{2}\ps\xi_0\ps\eta_0\ps\zeta_0\ps\pi_0\}$\\
\hspace*{2.4em}but that $\dd\star\ul{2}\ps\xi\ps\eta\ps\zeta\ps\pi$ is not.\\
\hspace*{2.4em}It follows from~(3) that \ $\dd\star\ul{2}\ps\xi\ps\eta\ps\zeta\ps\pi=\dd\star\ul{2}\ps\xi_0\ps\eta_0\ps\zeta_0\ps\pi_0$, i.e.\\ \hspace*{2.4em}$\xi=\xi_0,\eta=\eta_0,\zeta=\zeta_0$ and $\pi=\pi_0$. But this contradicts the hypothesis~:\\ \hspace*{2.4em}$\xi_0\star\pi_0,\,\eta_0\star\pi_0\notin\bbot_1^i$.

\qed

\begin{theorem} This realizability algebra is coherent.
\end{theorem}\noindent
Let \ $\theta\in$ QP be such that \ $\theta\star\pi^0\in\bbot_0^0$ and \ $\theta\star\pi^1\in\bbot_1^1$.
Then \ $\theta\star\pi^0\in\bbot_0^0\cap\bbot_1^0$ which contradicts lemma~\ref{bbot^i_0_1}.

\qed

\begin{lemma}
$\dd\,\ul{2}\force\;$(the boolean algebra $\gimel2$ has at most four $\veps$-elements).
\end{lemma}\noindent
We show that \ $\dd\,\ul{2}\force\pt x^{\gimel2}\pt y^{\gimel2}(x\ne0,y\ne1,x\ne y\to x\et y\ne x)$.\\
Let \ $i,j\in\{0,1\}$, $\xi\force i\ne0,\eta\force j\ne1,\zeta\force i\ne j$ \ and
$\pi\in\|i\et j\ne i\|$.\\
Since $\|i\et j\ne i\|\ne\vide$, we have $i\le j$. Thus, there are three possibilities for $(i,j)$~:\\
$i=j=0$~; $i=j=1$~; $i=0,j=1$.\\
In each case, two out of the terms $\xi,\eta,\zeta$ realize $\bot$.
Thus, we have \ $\dd\star\ul{2}\ps\xi\ps\eta\ps\zeta\ps\pi\in\bbot$.

\qed

\smallskip\noindent
{\small{\bfseries Remark}. If $\pi\in\Pi\setminus(\Pi^0\cup\Pi^1)$, then $\xi\star\pi\in\bbot$ for every term $\xi$.
Thus, we can remove these stacks and consider only $\Pi^0\cup\Pi^1$.}

\smallskip\noindent
We define two individuals in this realizability model~:\\
{$\gamma_0=(\{0\}\times\Pi^0)\cup(\{1\}\times\Pi^1)$~; $\gamma_1=(\{1\}\times\Pi^0)\cup(\{0\}\times\Pi^1)$}.\\
Obviously, $\gamma_0,\gamma_1\subset\gimel2=\{0,1\}\fois\Pi$. Now we have~:\\
$\|\pt x^{\gimel2}(x\neps\gamma_0)\|=\Pi^0\cup\Pi^1=\|\bot\|$ \ and therefore \
$\III\,\force\neg\pt x^{\gimel2}(x\neps\gamma_0)$.\\
$\dd\ul{0}\force0\neps\gamma_0$ and $\dd\ul{1}\force1\neps\gamma_0$.\\
It follows that $\gamma_0,\gamma_1$ are not $\veps$-empty and that every $\veps$-element
of $\gamma_0,\gamma_1$ is $\ne0,1$.\\
Therefore~:

\smallskip\noindent
{\em The Boolean algebra $\gimel2$ has exactly four $\veps$-elements.}

\smallskip\noindent
We have {$\xi\force\pt x^{\gimel2}(x\eps\gamma_0,x\eps\gamma_1\to\bot)$} for {\em every} term $\xi$~:\\
Indeed, let $i,j\in\{0,1\}$~; using lemma~\ref{nAtoB}, we replace the formula $i\eps\gamma_j$,
i.e. $\neg(i\neps\gamma_j)$, with $^{\neg}(i\neps\gamma_j)$ which is $\{\kk_\pi\;;\;\pi\in\Pi^{i+j}\}$.
Therefore, we have to check~:\\
$\rho_0\in\Pi^0,\rho_1\in\Pi^1\Fl\xi\star\kk_{\rho_0}\ps\kk_{\rho_1}\ps\pi\in\bbot$ \  which is clear.

\smallskip\noindent
In the same way, we get~:\\
$\lbd x\lbd y\lbd z\,z\force\pt x\pt y(x\eps\gamma_i,y\eps\gamma_i,x\ne y\to\bot)$.\\
It follows that $\gamma_0,\gamma_1$ are singletons and that their $\veps$-elements are the two atoms of $\gimel2$.

\subsection*{$\gimel2$ has four $\veps$-elements and $\gimel\kappa$ is countable}\noindent
We now apply to the algebra ${\cal A}$ the technique expounded in section~\ref{gkappaden}, in order to
make $\gimel\kappa$ countable~; this gives a realizability algebra ${\cal B}$.\\
In this case, we have \ $\kappa=\NN$, and therefore \ $\kappa_+={\cal P}(\kappa)=\RR$.

\smallskip\noindent
Now, there is an elementary formula which express that the Boolean algebra $\gimel2$ has four $\veps$-elements,
for instance~: \ $\ex x^{\gimel2}\{x\ne0,x\ne1\}\land\pt x^{\gimel2}\pt y^{\gimel2}(x\ne1,y\ne1,x\ne y\to xy=0)$.\\
Therefore, the realizability model ${\cal N}_{\cal B}$ realizes the following two formulas~:

\smallskip\noindent
($\gimel2$ has four $\veps$-elements)~; ($\gimel\kappa$ is countable)~;\\
and therefore also NEAC by theorem~\ref{gimel_k_DC}.

\smallskip\noindent
Let us denote by $i_0,i_1$ the two atoms of $\gimel2$~; thus, we have $i_1=1-i_0$.

\smallskip\noindent
We suppose that ${\cal M}\models V=L$~; thus, there exists on $\kappa_+={\cal P}(\NN)=\RR$ a strict
well ordering $\trgl$ of type $\aleph_1$. This gives a function from $\RR^2$ into $\{0,1\}$, denoted
by $(x\trgl y)$, which is defined as follows~: \
$(x\trgl y)=1$ $\Dbfl$ $x\trgl y$.\\
We can extend it to ${\cal N}_{\cal A}$ and ${\cal N}_{\cal B}$, which gives a function from $(\gimel\RR)^2$
into $\gimel2$.\\
From lemmas~\ref{k+iprod} and~\ref{k+ibf}, we get~:

\smallskip\noindent
For $i=i_0$ or $i_1$, the relation \ $(x\trgl y)=i$ is a strict total ordering on $\gimel_i\RR$
and one of these two relations is a well ordering~;\\
in order to fix the ideas, we shall suppose that it is for $i=i_0$.\\
The relation \ $(x\trgl y)=1$ is a strict order relation on $\gimel\RR$, which is well founded.\\
The application $x\mapsto(i_0x,i_1x)$  from $\gimel\RR$ onto
${\gimel_{i_0}\RR\fois\gimel_{i_1}\RR}$ is an isomorphism of strictly ordered sets.

\smallskip\noindent
It follows from theorem~\ref{GEinfini}, that each of the sets $\gimel_{i_0}\RR$, $\gimel_{i_1}\RR$ contain
a countable subset.\\
By corollary~\ref{no_surj}, there is no surjection from each one of the sets
$\gimel_{i_0}\RR,\gimel_{i_1}\RR$ onto the other. Thus, there is no surjection from $\NN$
onto $\gimel_{i_0}\RR$ or onto $\gimel_{i_1}\RR$.\\
Therefore, the well ordering on  $\gimel_{i_0}\RR$ has, at least, the order type~$\aleph_1$ in ${\cal N}_{\cal B}$.

\smallskip\noindent
Now, by theorem~\ref{segdentrgl}, every subset of $\gimel\RR$, which is bounded from above
for the ordering $\trgl$, is countable~; thus, the same goes for the proper initial
segments of $\gimel_{i_0}\RR$ and $\gimel_{i_1}\RR$, since these sets are totally ordered
and $\gimel\RR$ is isomorphic to ${\gimel_{i_0}\RR\fois\gimel_{i_1}\RR}$.

\smallskip\noindent
It follows that the well ordering on $\gimel_{i_0}\RR$ is at most $\aleph_1$, and therefore exactly $\aleph_1$.\\
Moreover, \emph{there exists, on $\gimel_{i_1}\RR$, a total  ordering, every proper initial segment of which
is countable}.

\smallskip\noindent
Then, we can apply theorem~\ref{gk_den}, to the sets $X_{i_0},X_{i_1}$ which are the images
of $\gimel_{i_0}\RR,\gimel_{i_1}\RR$  by the injection from \ $\gimel\kappa_+$ into \ $\RR$, which is
given by theorem~\ref{gk+R}. By setting $X=X_{i_1}$, we obtain exactly the result (ii) of relative consistency announced in the introduction.

\end{document}